\documentclass[preprint2,times,tighten]{aastex6}

\shorttitle{New debris disks in nearby young moving groups}

\shortauthors{Mo\'or et al.}

\begin{document}


\title{New debris disks in nearby young moving 
groups}

\author{A. Mo\'or\altaffilmark{1}}
\email{moor@konkoly.hu}
\author{\'A. K\'osp\'al\altaffilmark{1,2}}
\author{P. \'Abrah\'am\altaffilmark{1}}
\author{Z. Balog\altaffilmark{2}}
\author{T. Csengeri\altaffilmark{3}}
\author{Th.~Henning\altaffilmark{2}}
\author{A. Juh\'asz\altaffilmark{4}}
\author{Cs. Kiss\altaffilmark{1}}

\altaffiltext{1}{Konkoly Observatory, Research Centre for Astronomy
  and Earth Sciences, Hungarian Academy of Sciences, P.O. Box 67, H-1525
  Budapest, Hungary}

\altaffiltext{2}{Max-Planck-Institut f\"ur Astronomie, K\"onigstuhl
  17, 69117 Heidelberg, Germany}
  
\altaffiltext{3}{Max-Planck-Institut f\"ur Radioastronomie, Auf dem
  H\"ugel 69, 53121 Bonn, Germany} 

\altaffiltext{4}{Institute of Astronomy, Madingley Road, Cambridge CB3, OHA, UK}


\begin{abstract}
A significant fraction of  nearby young moving group members harbor
circumstellar debris dust disks. Due to their proximity and youth, these disks
are attractive targets for studying the early evolution of debris dust and
planetesimal belts. Here we present 70 and 160{\micron} observations of 31
systems in the $\beta$\,Pic  moving group, and in the Tucana-Horologium,
Columba, Carina and Argus associations,  using the Herschel Space Observatory.
None of these stars were observed  at far-infrared wavelengths before. Our
Herschel measurements were  complemented by photometry from the WISE satellite
for the whole sample, and by submillimeter/millimeter continuum data for one
source, HD\,48370. We identified six stars with infrared excess, four of them
are new discoveries. By combining our new findings with results from the
literature, we examined the incidence and general characteristics of debris
disks around Sun-like members of the selected groups. With their 
dust temperatures of $<$45\,K the newly identified disks around HD\,38397,
HD\,48370, HD\,160305, and BD-20\,951 represent the coldest population within
this sample. For HD\,38397 and HD\,48370, the emission is resolved in the
70{\micron} PACS images, the estimated radius of these disks is $\sim$90\,au.
Together with the well-known disk around HD\,61005, these three systems
represent the highest mass end of the known debris disk population around young
G-type members of the selected groups. In terms of dust content, they resemble
the hypothesized debris disk of the ancient Solar System.
\end{abstract}






\section{Introduction}
\label{intro}


The extensive surveys of the last decades have achieved a fundamental progress in investigating the young 
stellar population in the immediate ($<$100\,pc) vicinity of the Sun. A significant fraction of the identified 
young stars belong to gravitationally unbound stellar associations, or {\it young moving groups}, 
whose members share common space motions and are believed to have a common origin. With their ages ranging 
from eight to 
a few hundred Myr, these groups are ideal targets to study the early evolution of stars and their circumstellar 
environment. The comparison of a sequence of well-dated groups can outline general evolutionary trends, 
while the dispersion of properties within coeval groups could reveal additional secondary effects. The covered 
time period also overlaps with the epochs of several key events in the early Solar System history \citep{apai2009}, 
and is thought to be 
an active phase in the evolution of planetary systems in general.

A noteworthy fraction of moving group members exhibits excess emission at infrared (IR) wavelengths \citep[e.g.][]{zuckerman2011}. Though
a few long lived primordial disks may be present around the youngest stars of these associations
\citep[e.g.][]{calvet2002,kastner2008}, in most cases the 
observed excess  
can be attributed to thermal emission of tenuous circumstellar debris 
disks that are composed of subplanetary sized bodies down to micron-sized 
dust grains. In such debris disks the small dust particles are effectively 
removed from the system in a short timescale due to their interaction with the stellar 
radiation and/or wind. However, erosion of larger planetesimals -- 
particularly through their mutual collisions -- provide a continuous replenishment of grains \citep{wyatt2008}. 

While the population of planetesimals is invisible for us, 
their presence, location and some characteristics 
can be deduced from observations of the dust grains as tracers. 
Due to this close relationship, the study of debris dust evolution can give insight into 
the evolution of planetesimal belts as well.
Moreover, in such young systems the dust production is not limited to planetesimal belts. 
The final accumulation of terrestrial planets -- thought 
to occur via giant embryo-embryo collisions -- can also release a large amount 
of smaller fragments in the inner regions \citep{jackson2012,genda2015}. In our Solar System 
these processes lasted $\sim$100 Myr \citep{jacobson2014}. Based on 
geochemical arguments the formation of Earth was completed in 50-100\,Myr
\citep{konig2011}. 
Giant planets start their lives warm and then cool off on a timescale of a few ten million years, 
 making their detection possible via direct imaging at near-infrared 
wavelengths around members of young moving groups. Detection of such planet(s) and a debris disk 
in the same system provides an excellent opportunity to study the possible planet-disk interactions 
as already demonstrated, e.g., in the $\beta$\,Pic system 
\citep[e.g.][]{beust1996,mouillet1997,apai2015}.

The ultimate aim of our program is to study the early evolution of  planetesimal belts via observing 
and comparing debris dust disks in young moving groups at different ages. This requires, however, a 
census of debris disks in these groups. Thus, as the first step of the present work, we check the 
inventory of disks in selected moving groups and complete their disk detection lists.
We searched for new debris disks with the {\sl Herschel Space 
Observatory} \citep{pilbratt2010} in the following five nearby young 
moving groups: the $\beta$ Pictoris Moving Group 
\citep[BPMG, $\sim$20\,Myr,][]{barrado1999,zuckerman2001,binks2014,mamajek2014}, 
the Tucana-Horologium \citep[THA, $\sim$20--50\,Myr,][]{zuckerman2000,torres2000,torres2008,kraus2014,bell2015}, 
the Columba and the Carina associations \citep[COL and CAR, 
$\sim$20--50\,Myr,][]{torres2008,gagne2014,bell2015}, and the recently 
identified Argus association \citep[ARG, $\sim$30--50\,Myr,][]{torres2003,torres2008,gagne2014}. 
Ages of these groups range between 20 and 50\,Myr, thus they constitute
an ideal time sequence for following the evolution of planetesimal belts after 
the dispersal of gas rich protoplanetary disks. 

Due to its limited 
sensitivity, the Infrared Astronomical Satellite {\sl IRAS} could detect only the brightest 
circumstellar disks in these groups. Using spectroscopic and photometric 
{data} obtained with the {\sl Spitzer Space Telescope} \citep{werner2004} 
 many additional members 
 {were studied} at mid- and far-infrared wavelengths.
  {\citet{rebull2008} present infrared photometry at 24 and 70{\micron} obtained with the {\sl Spitzer} MIPS 
detector for 30 BPMG and 9 THA members. 
Using the same instrument and bands, in an extended photometric survey \citet{zuckerman2011} observed 53, 8, and 7 members of 
the THA, COL, and ARG groups, respectively. Both \citet{rebull2008} and \citet{zuckerman2011} concluded that the typical 
disk incidence in these groups is higher than among older field stars.
}  
With the advent of the {\sl Herschel Space Observatory} \citep{pilbratt2010} 
it became possible to measure the far-IR emission of debris disks 
with an unprecedented sensitivity and spatial resolution, enabling the 
study of cold dust in the outer disk regions. The spectroscopic capability 
of observing lines such as \ion{O}{1} and \ion{C}{2} offered an opportunity 
to search for the possible gas content of the disks.  
{Using the photometer unit of the Photodetector Array Camera and Spectrograph 
\citep[PACS,][]{poglitsch2010}, \citet{donaldson2012} studied 17 THA members, while 
\citet{rm2014} observed 19 systems belonging to the BPMG.
These observations} resulted in identifying new debris systems, {as well as} a 
better characterization of some 
previously known ones, 
spatially resolved images of some disks 
\citep[e.g. 49\,Ceti,][]{roberge2013}, 
and gas detection in disks around two BPMG members 
\citep[][]{rm2012,rm2014}.  
  
However, the member lists of BPMG and THA were incomplete at the 
time of the {\sl Spitzer} cryogenic mission, while the discoveries of 
the COL, CAR, and ARG groups happened in the last period of the mission 
excluding their systematic studies with {\sl Spitzer}. Observations with 
{\sl Herschel} \citep{donaldson2012,rm2014} were mainly focused 
on those well-known members of the groups that were already observed by
{\sl Spitzer}. Thus for most of the recently identified members 
of the specific  groups only mid-IR data are available from the 
WISE all-sky survey.
In our program we searched for far-IR excess over the stellar photosphere using {\sl Herschel} for  
those 
overlooked members of the above mentioned five groups that have never 
been observed at wavelengths longer than 25{\micron}.
In one case we also performed IRAM observations, in order 
to further characterize the newly discovered disk.

\section{Observations and data reduction} \label{obsanddatared}

\subsection{Sample selection} \label{sample}
To assemble our sample we compiled member lists of the five selected young moving groups 
based on the literature at that time \citep{torres2008,dasilva2009,zuckerman2011},
and on our previous surveys 
\citep{kiss2011,moor2013}. 
From these lists we selected those B9-K7 type stars, that have never been observed 
by {\sl Spitzer} or {\sl Herschel} at far-IR wavelengths. We focused predominantly on stars located 
within 80\,pc. 
Out of the 42 selected stellar systems, 26 could be observed before the end of the {\sl Herschel} mission. 
Because two targets were wide separation multiple systems, altogether
we obtained {\sl Herschel} observations for 28 stars. For three additional young moving group 
members, far-IR photometric data were taken from the Herschel Science Archive. Thus altogether 
31 objects were studied in this project.
The observed sample is dominated by G- and K-type stars, allowing us also to search for potential 
analogs of the young Solar System's debris disk.
Table~\ref{stellarprops} lists the main properties of the selected targets.


\subsection{Observations with the {\sl Herschel Space Observatory}}  \label{pacsdata}

The photometer unit of the Photodetector Array Camera and Spectrograph 
\citep[PACS,][]{poglitsch2010} was used 
to obtain far-IR measurements for the targets. PACS was a dual band 
imaging camera that provided 
simultaneous observations in two bands at 70/160\,{\micron} or at 100/160\,{\micron}.
Imaging 
was performed by two bolometer detector arrays with 64$\times$32 pixels (blue array) in the 70 and 100\,{\micron} bands and 
with 32$\times$16 pixels (red array) in the 160\,{\micron} band. 
The observations were carried out at 70/160\,{\micron} in mini scan-map mode, which is designed for photometry of 
individual point sources.
The maps were executed with a scan speed of 20{\arcsec}\,s$^{-1}$, using 10 scan-legs of 3{\arcmin} length 
separated by 4{\arcsec}.  All targets were observed at two scan angles of 70{$\degr$} and 110{$\degr$}. 
Table~\ref{obslog} presents the log of the {\sl Herschel} observations.

\paragraph{Data processing of PACS maps.} 
The data reduction was carried out in the Herschel Interactive Processing Environment 
\citep[HIPE,][]{ott2010} version 13 using PACS calibration tree No.\,65 and 
the standard HIPE script optimized for mini scan-map observations. 
We used only those data frames from the timeline where the actual scan speed of the spacecraft 
was between 15 and 25{\arcsec}~s$^{-1}$. 
The low-frequency ($1/f$) noise of the detector 
was eliminated by applying a highpass filtering with filter width parameters of 20 and 35 
(corresponding to 82{\arcsec}, and 142{\arcsec}), for
the 70 and 160{\micron} data, respectively.  
{In the determination of the highpass filter widths we basically followed the outline of \citet{balog2014}, except that at
70{\micron} we increased the filter width parameter from 15 \citep[corresponding to point sources]{balog2014} to 20 thereby considering the
possible presence of sources with up to $\sim$5{\arcsec} spatial extent in the sample.}
To avoid flux loss caused by this process, the immediate vicinity of our targets 
was excluded from the filtering using a 25{\arcsec} radius circular mask placed at the sources' position.
Glitches were removed using the second-level deglitching algorithm. 
Finally we used the {\sc photProject} task to combine all data frames belonging to a source 
into maps with pixel sizes of 1\farcs1 and 
2\farcs1 at 70 and 160\,{\micron}, respectively.


The red PACS detector was assembled from 2 bolometer matrices with 16$\times$16 
pixels each. After operational day 1375 (2013 February 16) one of the two red channel subarrays 
was affected by a serious anomaly and therefore became unusable. 
The failure of this matrix caused a small reduction in spatial coverage of 160\,{\micron} maps 
compared to shorter wavelength maps made with the blue detector array, 
and accompanied a loss of $\sim$30\% in sensitivity in this channel 
(see Herschel Observers' Manual version 5.0.3\footnote{http://herschel.esac.esa.int/Docs/Herschel/html/Observatory.html}). 
We note that the other 16$\times$16 pixel subarray remained stable. By studying dedicated observations of 
PACS internal calibration sources throughout the mission it was demonstrated that 
the measured signal level
obtained with this part of the detector was not changed after this event \citep{moor2014}.
Apart from the cases of HD\,38397, HD\,160305, and CD-54 7336,
all of our 160\,{\micron} observations are affected by this issue.

\paragraph{PACS photometry.}

At 70\,{\micron} nine of our 31 targets were detected at $\geq 3\sigma$ level.
From these nine sources four (HD\,160305, HD\,38397, HD\,48370, and BD-20 951) were detected 
in the 160\,{\micron} band as well.
The positional offsets between the centroids of the identified point sources 
and their 2MASS positions (corrected for the proper motion between the epochs of the two observations) 
are less than 1\farcs5 for all of our sources, thus the offsets
are within the 2$\sigma$ uncertainty of {\sl Herschel}'s 
nominal pointing accuracy \citep[$\sim$1\farcs0,][]{sportal2014}. 

We used aperture photometry for all of our targets. For the detected sources the 
aperture was placed around the centroid of the object,
while in the undetected cases the sources' proper motion-corrected 2MASS positions
were used as the target coordinates. 
Generally the aperture radii were set to 8{\arcsec} and 12{\arcsec} at 70 and 160\,{\micron}, respectively, while 
the sky background was estimated in an annulus between 40{\arcsec} and 50{\arcsec}.
In three special cases we deviated from these standard aperture sizes.
To avoid contamination, for HD\,35996 and HIP\,25434, which constitute a binary with a separation of 12{\arcsec},  
we used smaller apertures with radii of 6{\arcsec} and 8{\arcsec} in the 70 and 160\,{\micron} bands, respectively. 
In the case of HD\,38397 and HD\,48370, that turned out to be spatially extended 
(Sect.~\ref{modelling}), 
we used 
15{\arcsec} radius apertures in both bands. 
 In order to account for the flux outside the aperture we applied aperture correction, by using correction 
factors taken from the appropriate calibration files.
To compute the sky noise, we distributed sixteen apertures with radii identical to the
source aperture evenly along
the background annulus and performed aperture photometry without background subtraction 
in each of them, determining the noise as the standard deviation 
of these background flux values.  
The final photometric uncertainty was calculated as a quadratic sum of 
the sky noise and the absolute calibrational uncertainty of the PACS detector 
\citep[7\%,][]{balog2014}. The resulting 70 and 160\,{\micron} photometry is presented in 
Table~\ref{phottable}.



\paragraph{Serendipitious SPIRE/PACS parallel maps for HD\,48370.}
One source, HD\,48370, that turned out to exhibit the brightest infrared excess 
in our sample (Sect.~\ref{modelling}), was serendipitously measured in four observations 
(OBSIDs: 1342250790, 1342250791, 1342253429, 1342253430) in the SPIRE/PACS parallel 
mode (Herschel Observers' Manual version 5.0.3). In this mode, the SPIRE and PACS cameras were operated in photometry 
mode simultaneously generally performing large-area mapping observations with a scan speed of 60{\arcsec}/s. 
In all four 
cases the PACS camera was used with the 70\,{\micron} filter, thereby these observations 
provide data at 70, 160, 250, 350, and 500\,{\micron}.   

The processing of the PACS 70 and 160\,{\micron} data was done with 
HIPE utilizing the same reduction steps 
as described above, except that because of the three times 
higher scan map speed, the data frames were limited to those where the actual scan speed 
was between 48 and 72{\arcsec}~s$^{-1}$, and we set smaller highpass filter 
width parameters of 7 and 12, at 70 and 160\,{\micron}, respectively.
As a final step, a small region of 4{\arcmin}$\times$4{\arcmin} centered on HD\,48370 
was extracted from all four maps and combined in 
both bands. 
  For the photometry we used the same aperture setup as in the case of the mini-map observation of  
the source (see above). 
The resulting fluxes are listed in Table~\ref{phottable}.
In both bands these fluxes are in agreement with the ones 
obtained from the mini-map observations above (see Table~\ref{phottable}), 
within formal uncertainties. 
Thus for further analysis we combined the fluxes in each band 
using their weighted average (Table~\ref{phottable2}). 

Apart from one observation (1342253430), where the source was located very close to 
the edge of the map, HD\,48370 was detected in all bands in the other three SPIRE data sets 
(OBSIDs: 1342253429, 1342250790, 1342250791).
We applied the {\sc Timeline Fitter} task in HIPE to derive SPIRE photometry.
This task fits two-dimensional elliptical or circular 
Gaussian functions to the baseline-subtracted timeline data
at the coordinates of the source \citep[for details, see][]{bendo2013}.
We used circular Gaussians and the default settings for both the 
search radii (22, 30, and 42{\arcsec} in 250, 350, and 500\,{\micron} bands) 
and the background annulus (between 300 and 350{\arcsec} centred 
on the source). As a final step, the individual flux density values  
derived in the three maps in each band were averaged, weighted by their 
uncertainties. 
The final uncertainties
were derived as the quadratic sum of the measurement errors and the
5.5\% overall calibration uncertainty of the SPIRE photometer \citep{bendo2013}.
This method yielded flux densities of 148.5$\pm$12.1\,mJy, 
103.4$\pm$11.6\,mJy, and 54.5$\pm$10.5\,mJy at 250, 350, and 500\,{\micron}, 
respectively (see also Table~\ref{phottable2}).

\subsection{IRAM/NIKA observation}
After discovering it with Herschel, we initiated millimeter wavelength observations of HD\,48370
using the New IRAM KID Array (NIKA) dual-band imaging camera mounted on the 
IRAM 30-m telescope at Pico Veleta. 
NIKA is equipped with two novel type Kinetic Inductance Detector arrays of 132
and 224 detectors allowing simultaneous observations in two bands
at 1.25\,mm and 2.14\,mm, respectively, with a common field-of-view of 1$\farcm$9
in diameter \citep{catalano2014}. 
The observation was performed on 2014 February 25 in the framework of the first NIKA 
observing pool session. 
Pointing measurements and corrections were performed every 90 minutes, while 
bright sources were observed in every two to three hours for checking the focus 
of the telescope. 
Uranus was used for absolute flux calibration.
In order to map the vicinity of our target, 
we used the Compact Source observing template and executed 12 consecutive 1$\farcm$5$\times$1$\farcm$5   
Lissajous maps centered on the source. The total integration time was 1\,hr. 
During the observation the mean opacities were 0.174 and 0.139 at 1.25 and 2.14\,mm, respectively.
For processing our observations we used the NIKA data analysis pipeline (version 1). 
The details of the main data reduction steps performed in the pipeline 
are described in \citet{catalano2014}. The calibration was done assuming a 
Gaussian main beam of 12$\farcs$5 and 18$\farcs$5 (FWHM) at 1.25 and 2.14\,mm, respectively.
Besides flux calibrated maps for the 12 individual observations, the pipeline also produced 
a final combined map in each band.
HD\,48370 was clearly detected in both bands. Its shape is consistent with a 
point-like source. 
The pipeline provided photometry for the central source in each individual maps.
The final flux densities and their uncertainties at both wavelengths 
were computed 
as a weighted average of the 12 individual flux estimates. 
 According to \citet{catalano2014} the overall calibrational uncertainty is 15\% in the shorter and 
10\% in the longer wavelength band. Considering these uncertainties
by adding them quadratically to the measurement errors, our 
observation yielded flux densities of 7.3$\pm$1.5\,mJy and 2.0$\pm$0.3\,mJy
 at 1.25 and 2.14\,mm, respectively (Table~\ref{phottable2}).  

\section{Results} \label{dataanalysis}

\subsection{Stellar properties} \label{stellarprop}
For characterization of {the infrared excess related to circumstellar dust},
one has to compare the predicted 
photospheric flux densities of the star with the observed ones.
We modeled the stellar photospheres by fitting an {\sc ATLAS9} 
atmosphere model \citep{castelli2004} to the optical and infrared photometric data 
of the given target.
Optical photometry were taken from the {\sl Hipparcos} and {\sl TYCHO2} catalogs 
\citep{perryman1997,hog2000}, and from the  
SACY survey \citep[Search for Associations Containing Young stars,][]{torres2006}. 
{Zero-point flux densities for optical bands were taken from the SVO Filter Profile 
Service\footnote{http://svo2.cab.inta-csic.es/svo/theory/fps3/}.}
At near-IR wavelengths we used the $J, H, K_s$ data measured in the Two Micron All-Sky Survey \citep[2MASS;][]{skrutskie}.
All of our targets have counterparts in the Wide-field Infrared Survey Explorer \citep[{\sl WISE},][]{wright2010} 
catalog that provides mid-IR photometry in four bands, $W1, W2, W3$ and $W4$ centred at 3.4, 4.6, 12, and 22{\micron}, 
respectively. In case the confusion flag indicates an unconfused measurement 
in the $W1$ band (i.e. cc\_flg[$W1$] = 0), these data were also added to the optical and near-IR 
photometry. 
{For 2MASS and WISE $W1$ data the relevant zero-point values were taken 
from \citet{cohen2003} and \citet{wright2010}, respectively.}
We adopted solar metallicity for all of our sources and, except for HD\,32309, where we 
assumed a $\log{g}$ of 4.0 (cgs units), the surface gravity was fixed to $\log{g}=$4.5.
Apart from CD-52 1363 and CD-42 2906, the selected stars are located within 80\,pc, in the cavity 
known as the Local Bubble, where the mean extinction is expected to be low \citep[e.g.,][]{lallement2003}.
For the two more distant objects we collected all those stars in their 3{\degr} vicinity 
that are included in the {\sl Hipparcos} catalog and having Str\"omgren color indices and H$\beta$ index 
in the photometric catalog of \citet{hauck1998}.
By deriving $E(B-V)$ color excesses (as $E(B-V) = E(b-y) / 0.74$) from the Str\"omgren photometry using 
the appropriate calibration processes \citep{crawford1975,crawford1979,olsen1984} we found 
that the reddening of stars located within 150\,pc are $<$0.015\,mag. 
Considering these results, the visual extinction was neglected for all targets in the fitting process.
A grid-based approach was used in the fitting. 
The effective temperature ($T_{\rm eff}$) values of the best fitting models as well as 
the derived stellar luminosities
are listed in Table~\ref{stellarprops}.

\subsection{Stars with infrared excess} \label{excess}
In order to identify systems exhibiting excess in any of the PACS bands, first we 
computed the predicted photospheric flux densities of the sample stars 
using their best-fit Kurucz models (Sect.~\ref{stellarprop}). 
The significance level of excess in a specific
photometric band was calculated as:
\begin{equation}
\chi_{\rm band} = \frac{F_{\rm band} - P_{\rm band}}{\sigma_{\rm band}^{\rm }}, 
\end{equation}
where $F_{\rm band}$ is the measured flux density either at 70 or 160\,{\micron}, 
$P_{\rm band}$ is the predicted stellar flux,
while $\sigma_{\rm band}$ is the 
quadratic sum of the final photometric uncertainty and the uncertainty of the predicted flux density 
in the relevant band. The accuracy of the predicted photospheric fluxes is estimated to be
around 5\%. It is important to note that in all cases the dominant source of excess error 
is the final photometric uncertainty.
The predicted photospheric flux densities ($P_{70}$, $P_{160}$) as well as 
the significances are listed in Table~\ref{phottable}.
Targets with $\chi_{\rm band} > 3$  in any of the PACS bands were considered 
as a star with excess emission.
Applying this criterion, we identified five stars, HD\,160305, HD\,38397, HD\,35996, HD\,48370, and 
BD-20 951 that exhibit excess emission at 70\,{\micron}. 
Apart from HD\,35996, all of these sources show excess at 160{\micron} as well. 
HD\,168210 has a $\chi_{\rm 160}$ of 3.7, however its photometry at 160{\micron} is strongly 
contaminated by a bright nebulosity coincident with its position, thus the observed excess may not come
from a circumstellar disk but rather
have interstellar origin.

WISE photometry
is available for all of our targets, enabling us to search for excess at
12 and 22{\micron} ($W3$ and $W4$ bands). 
The {\sl WISE} magnitudes and uncertainties  
were converted to flux densities using the formulae from
\citet{wright2010}. In the calculation of the final observational uncertainties 
the absolute calibrational 
errors\footnote{http://wise2.ipac.caltech.edu/docs/release/allsky/expsup/ \\ sec4\_4h.html\#CalibrationU} 
were also taken into account by quadratically adding to 
the listed measurement errors.
Then the significance level of the infrared excess was computed in the same way as 
for the PACS photometry. 
None of our targets have excess at 12{\micron}.
Four stars, HD\,38397, HD\,35996, HD\,48370, and CD-54 7336 
exhibit excess in the $W4$ band with significances of $\chi_{\rm W4} > 3$. 
Among them HD\,38397, HD\,35996, and HD\,48370 show excess in PACS bands as well, while 
CD-54 7336 has an excess only in the $W4$ band. 
Apart from HD\,35996, all of these sources have a good quality photometry in all WISE bands.
However, HD\,35996 is flagged as an extended source, 
because its measured source profiles in the $W1$ and $W2$ bands deviate significantly
from the WISE point spread function (i.e. the reduced $\chi^2$ 
of the $W1$ and $W2$ profile-fit photometry measurements are $>$3). 
According to the catalog's contamination and confusion flag 
the source photometry is contaminated by the scattered light halo surrounding a 
nearby bright source in these bands. Indeed HIP\,25434, the companion of HD\,35996, is located 
at a separation of $\sim$12{\arcsec}. The WISE catalog provides separated photometry for the 
two components in $W4$ band. No quality flags are assigned to these sources in this band and the 
measured flux of HIP\,25434 corresponds well to its predicted photospheric flux 
at the specific wavelength, suggesting that the observed excess at HD\,35996 might not be 
significantly contaminated by the companion. 
The WISE excess of HD\,38397 has already been reported by several authors 
\citep{wu2013,patel2014,vican2014}. Interestingly, according to \citet{patel2014}, 
HD\,160305 also exhibit excess at 22{\micron}, although for this star we derived an excess significance 
of $\chi_{\rm W4} = 2.8$.
 
\begin{figure*} 
\begin{center}
\includegraphics[scale=.43,angle=0]{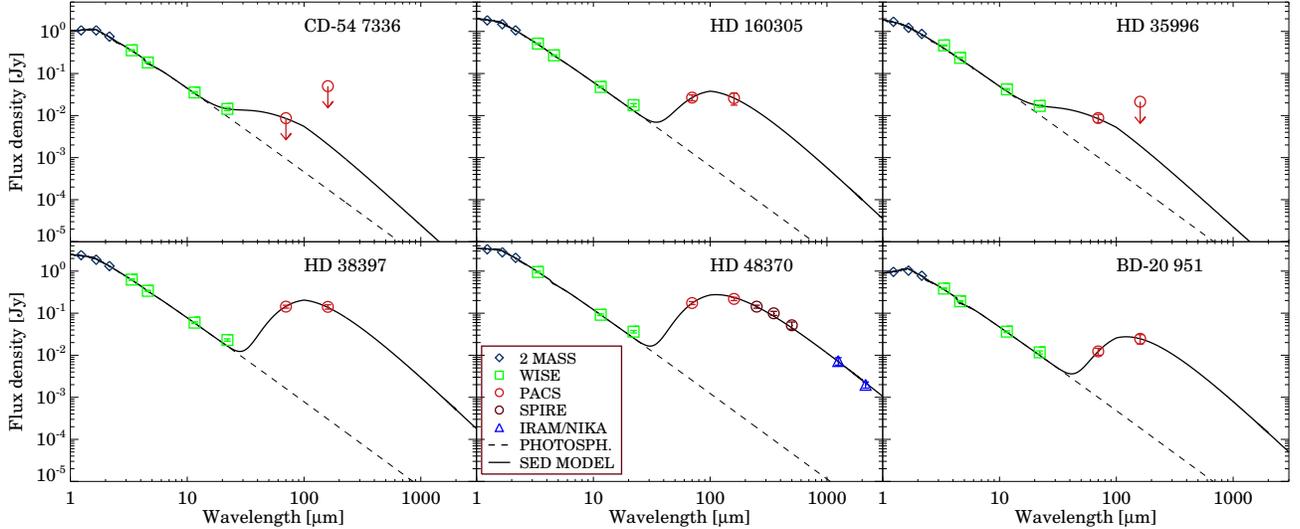}
\caption{ { Color corrected SEDs of stars with infrared excess.
}
\label{sedplot}
}
\end{center}
\end{figure*}

Altogether, we identified six stars with infrared excess, from which four, 
CD-54 7336, BD-20\,951, HD\,35996, and HD\,48370 are new discoveries. 
Their spectral energy distributions (SEDs), using 2MASS, WISE, and Herschel (and NIKA for HD\,48370) photometry, 
are shown in Figure~\ref{sedplot}.

\subsection{Disk properties}  \label{modelling}

\paragraph{Simple blackbody models.}
The infrared excess emission of main-sequence stars is generally linked to the 
thermal emission of an optically thin second-generation circumstellar 
dust disk heated by the central star. 
In most debris systems 
the observed excess can be well described 
by a single-temperature (modified) blackbody or a combination of two different 
temperature blackbody components \citep{kennedy2014,chen2014}. 
We adopted a single temperature modified blackbody model to estimate the characteristic 
dust temperature ($T_{\rm dust}$) and the fractional luminosity 
($f_{\rm dust} = \frac{L_{\rm IR}}{L_{\rm bol}}$) of the six systems with IR excess. 
In the case of CD-54 7336, where the excess was measured  only in one band we fitted 
the 3-$\sigma$ upper limit at 70{\micron} to constrain the dust temperature. 
The modified blackbody is needed to account for another general characteristic 
of debris disk SEDs: the steep fall of the excess 
spectrum at wavelengths significantly longer than the typical 
grain size because of the inefficient 
absorption and emission.
In our model the emissivity was set to 1 at $\lambda \leq \lambda_0$ wavelengths, 
while at longer wavelengths it varied as $(\lambda / \lambda_0)^{-\beta}$. 
Following \citet{williams2006} the $\lambda_0$ parameter was fixed to 100{\micron}. 
Apart from HD\,48370, whose millimeter SED is well characterized, for the other targets 
$\beta$ was fixed to 0.65  
\citep[][]{gaspar2012}.
We applied a Levenberg-Marquardt algorithm \citep{markwardt2009} to identify 
the best-fitting model to the measured data. Following  
\citet[][]{moor2006}, simultaneously with the fitting process we used 
an iterative method to compute
and apply color corrections for the photometric data. 
Table~\ref{diskparams} shows the fundamental disk parameters and their uncertainties derived from the
modeling. The fitted models are displayed in Fig.~\ref{sedplot}. 

Assuming that the blackbody components correspond to narrow dust belts that contain 
large grains emitting like blackbodies, based on the derived temperatures and the 
stellar luminosities (Table~\ref{stellarprops})
the rings' radii can be calculated as 
\begin{equation}
{R_{\rm bb}} [{\rm au}] = \left(\frac{L_{\rm star}}{L_{\sun}}\right)^{0.5} \left(\frac{278\,K}{T_{\rm dust}}\right)^2.
\end{equation} 
In reality, smaller grains --
that are ineffective emitters and thereby hotter than large blackbody particles at the same stellocentric distance -- 
may also be present in the disk, thus the derived $R_{\rm bb}$ values (Table~\ref{diskparams}) 
should be 
considered as minimum possible radii.
Indeed, the disk radii inferred from spatially resolved far-IR images of debris disks were found to be 
2--4 times larger than blackbody disk radii derived from the characteristic dust temperatures
\citep[e.g.][]{booth2013,morales2013,pawellek2014}.

Analyzing the SEDs of debris systems where good quality mid-IR spectra are also available, 
\citet{ballering2013} and \citet{chen2014} found that disks with two temperature components 
are common. Most of these two-temperature disks may be linked
to spatially distinct dust belts \citep{kennedy2014}.  
In the case of HD\,38397 and HD\,48370 (and to a lesser extent of HD\,160305 as well) 
the measured 22{\micron} flux density deviates significantly 
from the fitted single temperature model, hinting at a second temperature
component. However, without additional good quality mid-IR observations, 
this possible warmer component cannot be further characterized. 
The SED of HD\,48370, with its better wavelength coverage, allowed us to scrutinize how 
an additional component would modify 
the parameters of the cold one, by utilizing a two-component model where the warm component 
is represented by a simple blackbody.
By setting different temperatures for the warm 
blackbody and repeating the fitting procedure we found that if the warmer component 
has a temperature of $>$90\,K then the temperature and the $\beta$-parameter 
of the cold dust do not vary significantly compared to the single temperature model.   
For the fractional luminosity, however, we obtained values ranging between 6.5$\times$10$^{-4}$ 
and 7.3$\times$10$^{-4}$, which are higher than the one, $f_{\rm dust} \sim 5.6\times10^{-4}$, 
derived in the simpler model.

For those systems where no infrared excess has been identified
upper limits were computed for the fractional luminosities. 
Here we assumed a disk whose emission can be described by a modified blackbody 
(with $\beta=0.65$ and $\lambda_0=100${\micron}) that peaks at 70{\micron} 
with an amplitude corresponding to 3$\times\sigma_{70}$. 
The obtained upper limits are listed in Table~\ref{phottable}.

\paragraph{Analysis of millimeter data at HD\,48370.} 
Thanks to the SPIRE and IRAM/NIKA observations the SED of HD\,48370 is 
well characterized in the millimeter regime providing an opportunity to 
constrain the size distribution of the
emitting grains and estimate the dust mass.
For optically thin dust emission the measured flux density can be described 
as 
$F_{\nu, \rm dust} \propto \frac{M_{\rm dust} B_{\nu}(T_{\rm dust}) \kappa_{\nu}}{d^2},$ 
where $M_{\rm dust}$ is the dust mass, $B_{\nu}(T_{\rm dust})$ is the Planck function 
at the characteristic dust temperature, $\kappa_{\nu}$ is the dust opacity coefficient, while $d$ 
is the distance of the source. At long wavelengths the Planck function is proportional 
to $\nu^{\alpha_{\rm Planck}}$, while $\kappa_{\nu}$ can be given as $\kappa_{\nu} = \kappa_{0}
(\frac{\nu}{\nu_0})^{\beta}$. Considering these relationships and that 
$F_{\nu, \rm dust} \propto \nu^{\alpha_{\rm mm}}$, it means that 
$\alpha_{\rm mm} \approx \alpha_{\rm Planck} +\beta$.
In debris disks the size distribution of grains is generally described as 
$dN / da \propto a^{-q}$, where $a$ is the grain size,  
$N$ is the number of grains in a given size bin, while 
$q$ is the power-law index of differential size distribution of the dust grains. 
According to \citet{draine2006}, if the dust size distribution follows a power law between  
minimum and maximum grain sizes that are significantly 
smaller and larger than the observational wavelengths, respectively, and $3 < q < 4$, 
then the dust opacity power law index $\beta$ is related 
to $q$ as $\beta \approx (q-3)~\beta_s$, where $\beta_s$ is the dust opacity spectral 
index in the small particle limit. Thus $q$ can be estimated from the (sub)millimeter 
data using the following formula: 
\begin{equation}
q = \frac{(\alpha_{\rm mm} - \alpha_{\rm Planck})}{\beta_s} + 3,
\label{q} 
\end{equation} for $\beta_s$
we adopted a value of 1.8$\pm$0.2 \citep[taken from][]{ricci2012}.
Based on our color-corrected millimeter flux densities 
obtained between 0.35 and 2.14\,mm, the $\alpha_{\rm mm}$ parameter is 2.15$\pm$0.11, while 
 $\alpha_{\rm Planck}$ is 1.71 in the same wavelength range adopting a temperature of  
 41\,K (corresponding to the characteristic temperature of the disk). 
 Using these values Eq.~\ref{q} gives $q$=3.25$\pm$0.07. 

Millimeter observations also  enable us to estimate the dust mass in the 
disk. Assuming optically thin emission characterized by a single temperature 
the dust mass was computed as:
$M_{\rm dust} = \frac{F_{\nu, \rm dust} d^2}{ B_{\nu}(T_{\rm dust}) \kappa_{\nu}}$. 
By adopting 2\,cm$^2$\,g$^{-1}$ for $\kappa_{0}$ at $\nu_0 = 345$~GHz \citep[e.g.][]{nilsson2010} 
and using observations between 0.35 and 2.14\,mm we derived four individual estimates and then 
the final dust mass was computed as a weighted average of them, yielding $M_{\rm dust} = 0.12\pm0.04$\,M$_\oplus$.

\begin{figure*} 
\includegraphics[scale=.50,angle=0]{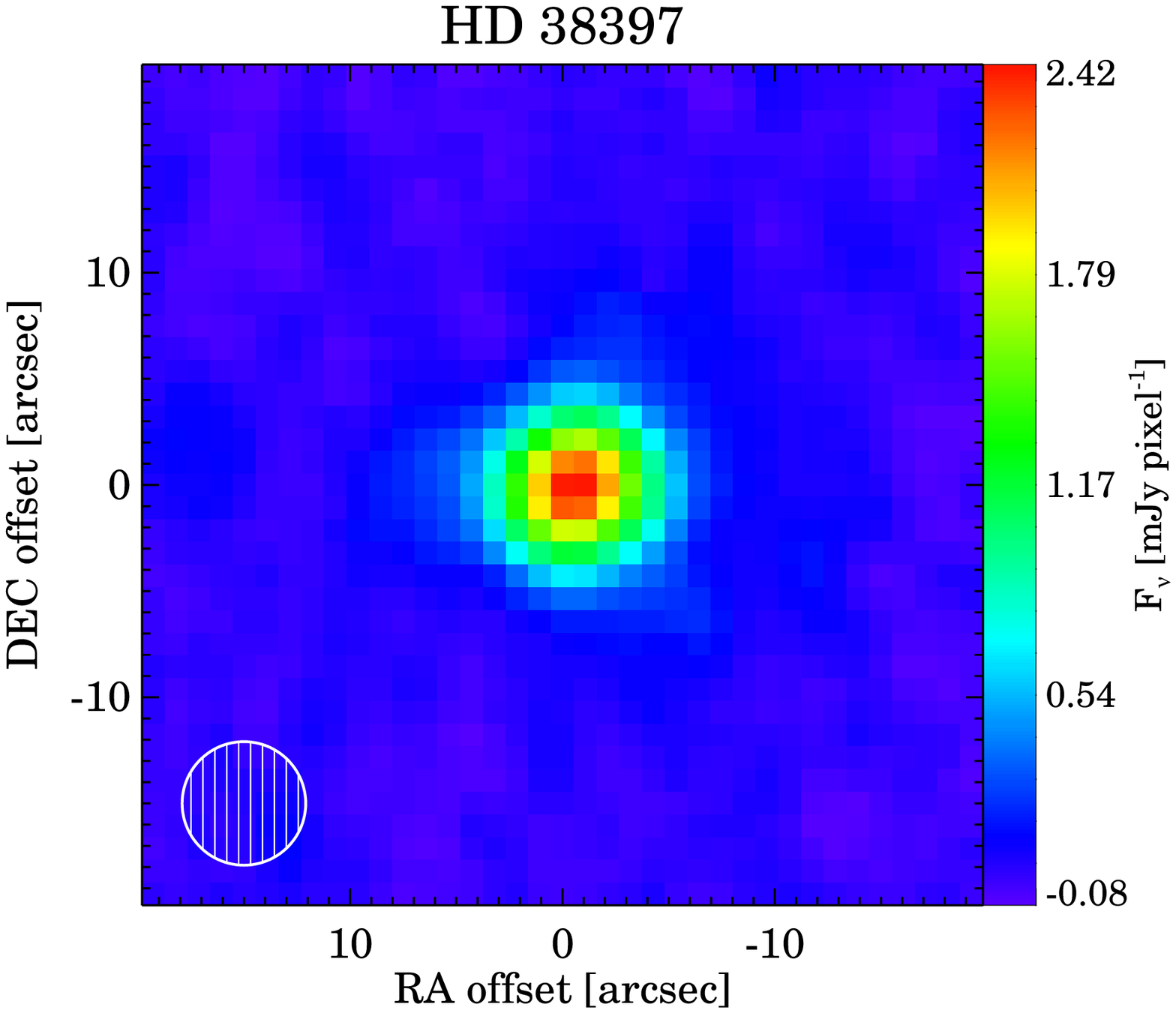}
\includegraphics[scale=.50,angle=0]{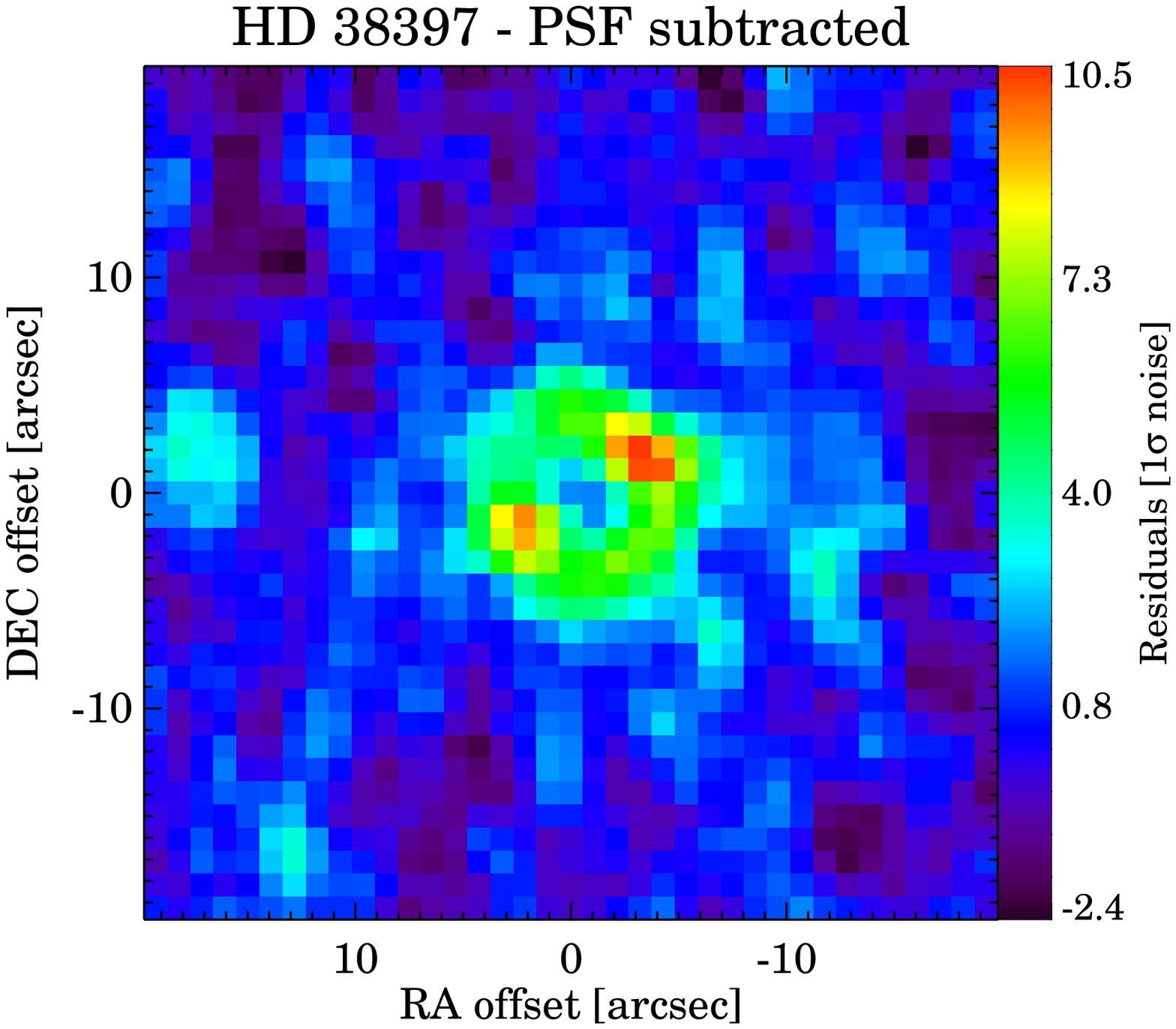}
\includegraphics[scale=.50,angle=0]{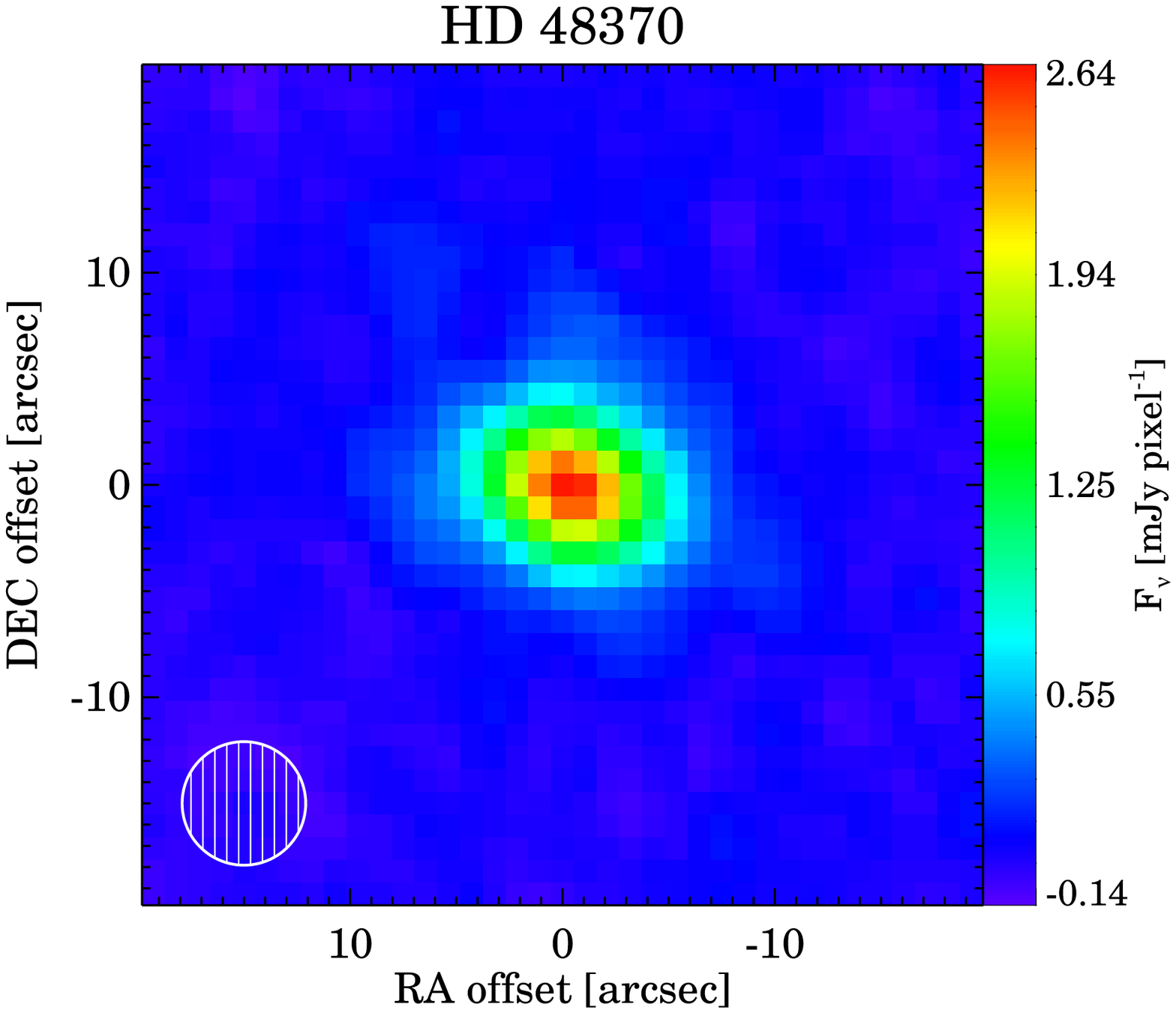}
\includegraphics[scale=.50,angle=0]{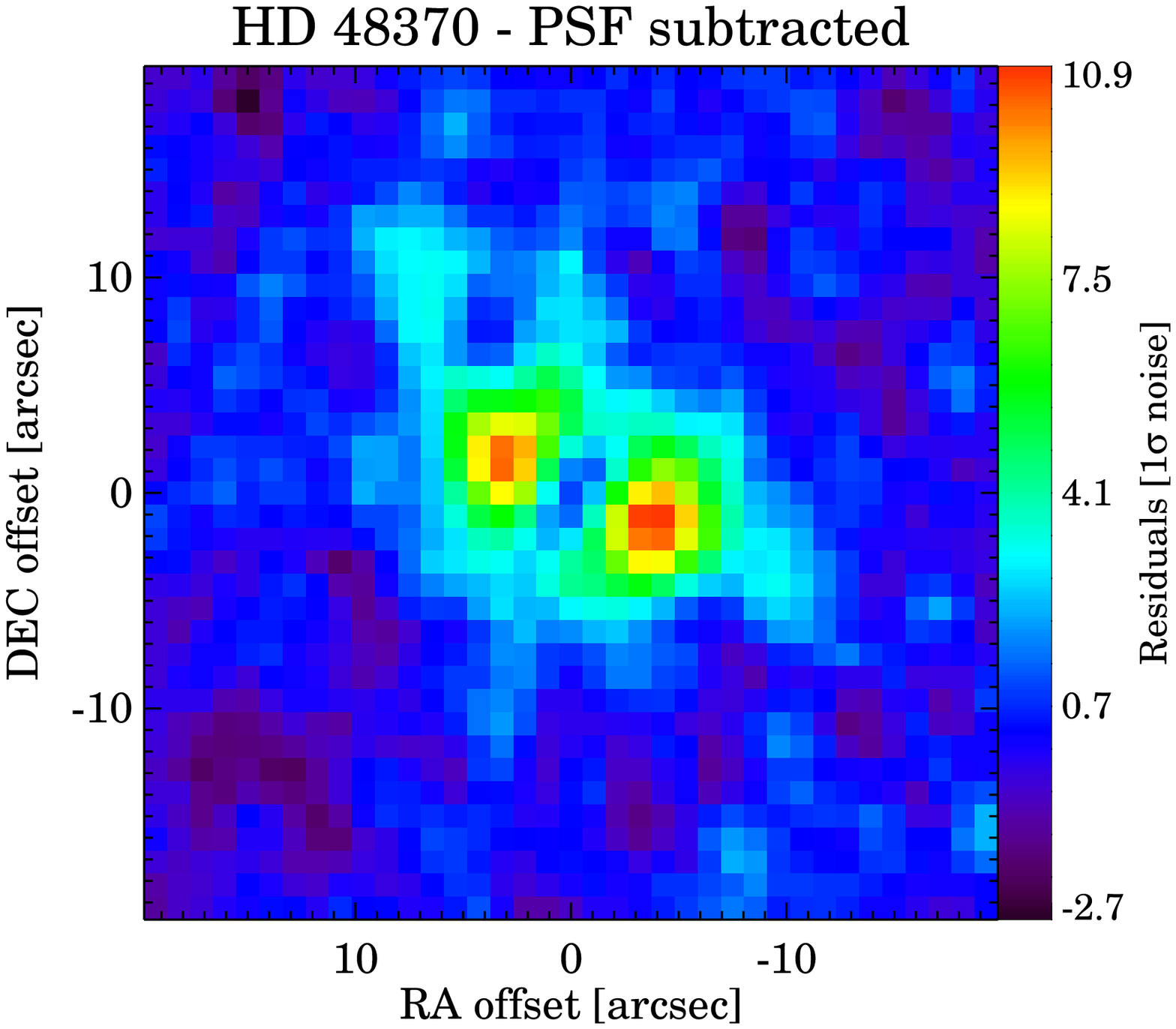}
\caption{ { Left: Herschel/PACS 70$\mu$m image of HD\,38397 and HD\,48370
(the hatched circle in the lower left corner shows the FWHM 
of the PSF). Right: PSF-subtracted 70$\mu$m image of HD\,38397 and HD\,48370.
}
\label{herschel}
}
\end{figure*}

\paragraph{Extended emission around HD\,38397 and HD\,48370.} 
For systems with excess in the PACS bands
we scrutinized whether the observed sources are point-like or spatially extended by comparing 
the measured source profiles with the appropriate point-spread functions (PSFs). 
To compile reference PSFs we used 70 and 160\,{\micron} mini-scan map observations of four diskless stars 
($\alpha$~Boo, $\alpha$~Tau, $\alpha$~Cet, and $\beta$~And) 
which served as photometric standards in the calibration of PACS \citep{balog2014}. 
These data were processed in the same way as
described in Sect.~\ref{pacsdata} and the obtained PSFs were rotated to match the telescope's 
roll angle at the observation of the given target. 
Then we fitted 2D Gaussians to these PSFs and the most extended one (with the largest FWHMs) 
were subtracted  -- after a scaling to the peak -- from the measured profiles of our targets.
At 160{\micron} all of our targets were found to be consistent with a point-like source. 
However, as Figure~\ref{herschel} (right) demonstrates, the PSF subtraction leaves significant
residual emission at 70{\micron} around HD\,38397 and HD\,48370, indicating that their disks 
are 
resolved at this wavelength.

Assuming that the bulk of the observed 70{\micron} emission originates 
from a narrow cold dust ring, the PACS images of HD\,38397 and HD\,48370  
were fitted by a simple, geometrical disk model. 
The model disks have three free parameters, the average radius ($R_{\rm avg}$), 
the position angle ($PA$) and the inclination ($i$). Following \citet{booth2013} and \citet{moor2015} 
the width of the disk was fixed to 0.1$R_{\rm avg}$ and the surface brightness was assumed to be 
 homogeneous. The contribution of the stellar photosphere and the possible warm component 
 was neglected in the model.
The compiled disk images were then convolved with the appropriately rotated PSF (see above).
We used a grid based approach in the course of fitting, and a Bayesian analysis 
was performed to select 
the best model and estimate the uncertainties of the derived disk parameters 
\citep[for more details, see][]{moor2015}. 
For HD\,38397 our best solution has 
$R_{\rm avg} = 90\pm21$\,au 
, $PA = 116^{+52}_{-47}{\degr}$, and $i = 38^{+32}_{-28}{\degr}$ (where the face-on orientation corresponds 
to $i = 0{\degr}$).
In the case of HD\,48370 the following disk parameters were obtained: 
$R_{\rm avg} = 89\pm18$\,au,  
$PA = 66\pm10 {\degr}$, and $i = 69^{+21}_{-20}${\degr}.
The ratios of the disk radii inferred from 
the 70{\micron} images to the blackbody disk radii (Table~\ref{diskparams}) are 
$\sim$2.0 and $\sim$2.3 for HD\,38397 and HD\,48370, respectively. 
Several authors \citep{booth2013,pawellek2014} noticed that there is a trend, where this ratio 
($\Gamma = R_{\rm disk} / R_{\rm bb}$) 
is increasing with the decreasing stellar luminosity, implying that grains around 
more luminous stars are generally larger.   
By placing our objects in fig.~4b of \citet{pawellek2014} -- that shows $\Gamma$ values 
as a function of stellar luminosity for 34 debris disks -- we found that 
they are broadly consistent with the currently outlined trend.

By relaxing our assumption of narrow rings we found that the measured profiles can equally 
be fitted with broader disk models that have smaller inner and larger outer radii than 
rings with 0.1$R_{\rm avg}$ width. 
However, position angles, inclinations, and even average disk radii inferred from these models were in 
reasonable accordance with those derived in our narrow disk scenario.

\subsection{Contamination by background galaxies or diffuse nebulosity?}
So far we assumed that the observed excesses are related to circumstellar material.
However at such long wavelengths the probability of contamination by background galaxies 
is not negligible, especially in the case of faint sources \citep[e.g.,][]{gaspar2014}. 
The brightness and spatial extent 
of HD\,38937 and HD\,48370 provide strong evidences for the circumstellar nature of the excess emission. 
In order to estimate the probability of galaxy confusion for the other disk candidates, 
we followed the outline 
described in \citet{sibthorpe2013} who used the results of Herschel-based deep extragalactic surveys to 
quantify the relevance of this contamination issue. 
In our calculation we used the confusion beam size definition proposed by \citet{condon1974}.
According to this work the effective beam solid angle of an elliptical Gaussian beam with half power 
axes of $\theta_1$ and $\theta_2$ can be computed as:
$\Omega_e = \frac{(\frac{1}{4} \pi \theta_1 \theta_2)} {(\gamma-1) \ln{2}},$ 
where $\gamma$ represents the slope of the differential distribution of sources. 
For a scan speed of 20{\arcsec}\,s$^{-1}$, $\theta_1 = 5\farcs46$ and $\theta_2 = 5\farcs76$ 
at 70{\micron} (PACS's Observers Manual\footnote{http://herschel.esac.esa.int/Docs/PACS/pdf/pacs\_om.pdf}),
while, based on \citet{berta2011}, the $\gamma$ parameter is $\sim$2 in the relevant flux range. 
Thus the effective beam solid angle and radius is 35.6\,arcsec$^2$ and $3\farcs36$, respectively. 
Considering the average 3$\sigma$ uncertainty of the 70{\micron} photometry (6.9\,mJy) as our survey 
limiting flux density, the probability of confusion by one or more background galaxies 
within a $3\farcs36$ radius 
of a target is found to be 0.27\%.  
Since our sample includes 31 objects this means that the probability that one of them is
being contaminated is 8\%, while the chance that two of them are affected by 
source confusion is only 0.34\%.

The far-IR SED of those faint galaxies that could be responsible for such contamination
can typically be fitted by a modified blackbody with temperatures ranging between 20 and 29\,K 
\citep{gaspar2014}. The characteristic temperature of the excess at HD\,35996 is 
significantly warmer (Table~\ref{diskparams}) what is typical for faint galaxies, implying that the chance of 
contamination is even lower. However, with its temperature of 33$\pm$5\,K, the disk candidate 
at BD-20\,951 would belong to the coldest population of known debris systems. 
{Among our disk candidates the excess SED of BD-20\,951 is the most 
compatible with that of galaxies.}

For CD-54 7336 the infrared excess was detected only in the $W4$ band. The PACS 160{\micron} image of this source 
shows extended bright nebulosity around the star, this diffuse emission can be recognized 
at 70{\micron} as well. Though its $W4$ band photometry is free of quality issues, considering the relatively 
coarse spatial resolution ($\sim$12{\arcsec}) we cannot completely exclude the possibility that this 
nebulous environment also contribute to the measured flux density at 22{\micron}.

\section{Discussion}
\label{discussion}

Our study of 31 young moving group members resulted in the identification of 
6 systems with infrared excess. Though we cannot completely exclude that some of them 
are due to source confusion or extended interstellar emission, in the following we 
suppose that the observed excesses are linked to circumstellar debris dust grains. 
Four among these six systems are known to be single stars, HD\,35996 resides 
in wide binary system with a minimum separation of $\sim$840\,au.
The latter object as well as BD-20\,951
were found to be spectroscopic binary (Table~\ref{stellarprops}). 
Considering their derived blackbody dust radii of 6 and 55\,au  
(Table~\ref{diskparams}) 
these systems may harbour circumbinary disks.

\subsection{Possible fainter debris disks?} \label{fainterdisks}

\begin{figure} 
\includegraphics[scale=.45,angle=0]{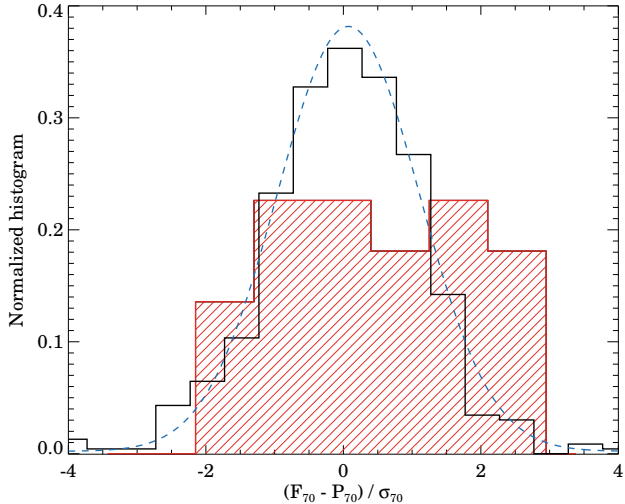}
\caption{ {Normalized histograms  of  the  significances  of  the  differences  between  the
measured and predicted photospheric flux densities at 70{\micron} for those of our targets 
where no excess has been identified (red, striped) and for photometric simulations performed 
in the vicinity of our targets (black histogram).  A
Gaussian fit to the histogram of simulated photometry provides a mean of
0.07 and a dispersion of 1.03 (blue dashed curve).   
}
\label{histo}
}
\end{figure}

The successfully observed disks have fractional luminosities of
$\gtrsim$10$^{-4}$ and thereby belong to the dustiest population among known debris disks. 
Actually,
this fractional luminosity value roughly corresponds to the detection limit for most 
of our targets, suggesting that several more tenuous disks could have remained undetected. 
Figure~\ref{histo} shows the normalized histogram of the significances of the differences between 
the measured and predicted  
flux densities at 70{\micron} for all those systems where no excess have been found.  
For comparison, we made a simulation, in which we performed aperture photometry 
in 100 randomly selected positions in the vicinity of each target. We used the same aperture 
size and background value as in the case of the targets, and the random apertures 
were placed between the target's aperture and the inner edge of the original background 
annulus. The normalized histogram of the 70{\micron} significances ($F_{\rm 70} / \sigma_{\rm 70}$ since 
in these cases $P_{\rm 70} = 0$) derived from these 
simulated photometry are also displayed in Figure~\ref{histo}.  
A Gaussian fit to this distribution yielded a mean of 0.07 and 
a dispersion of 1.03 in good agreement with the expected
mean of 0.0 with a dispersion of 1.0. The histogram of real measurements (blue, stripped) differs 
from this distribution by showing an excess at the positive side, suggesting 
that a fraction of our targets harbor undetected faint disks with lower fractional luminosities. 
 Indeed, some additional excess candidates can already be identified, e.g. for 
 HD\,51797, we measured $\chi_{\rm W4}$ and $\chi_{\rm 70}$ 
 significances of 2.4 and 2.7, respectively, while for HD\,32372  we obtained $\chi_{\rm 70} = 2.8$.  

\subsection{Debris disks around Sun-like members of young moving groups} \label{excessanalysis}
  
By merging our new results with the outcome of previous infrared surveys obtained either by {\sl Herschel} 
or {\sl Spitzer} for members of nearby moving groups, we 
investigated the general characteristics of debris disks around young Sun-like members.  
{Using member lists from the literature \citep{torres2008,moor2013,zuckerman2011,malo2013,bell2015} 
we collected all F8-K7 spectral type stars assigned 
to the five young groups studied in this paper.
We focused on objects located closer than 80\,pc, since beyond this distance the 
available far-IR data sets for the members are quite incomplete.
From the 103 F8-K7 type group members 67 have been observed at 70{\micron} (and some of them even at longer
wavelengths). 
One of these stars, V4046\,Sgr is a classical T\,Tauri binary system belonging to the BPMG, that
harbors a gas rich circumbinary disk. Therefore this object was discarded from the sample. 
From the remaining 66 sources,
25 were measured in our current study, for the other 41 sources 
information on the IR excess -- taken from the literature -- are summarized in Table~\ref{excesstable}.} 
{From the 66 systems 25 exhibit infrared excess, but only 
13 of them at 70{\micron}, the other objects have measurable excess only at shorter wavelengths.
It is reasonable to divide this sample into two subsamples, one with late F- and G-type stars (32 targets), 
and another one with K-type systems (34 objects). From the 32 F,G-type members 18 have  
 excess, 11 among them at 70{\micron} as well. In the subsample of K-type stars the detection rate 
 of debris disks is 7/34, at 70{\micron} two stars exhibit excess emission. These numbers demonstrate 
 that the detection rate in young K-type stars is lower than in earlier spectral classes.   
}
{As explained above (Sect.~\ref{fainterdisks}), the obtained 70{\micron} excess rates could be considered 
as lower limits. 
The 
detection rates at 70{\micron} in the five individual groups 
for the late F- and G-type stars are as follows: 
2/5 for the BPMG, 5/15 for the THA, 2/9 for the COL, 0/1 for the CAR, and 2/2 for the ARG. 
The same rates for the K-type stars: 1/8 in the BPMG, 0/12 in the THA, 0/7 in the COL, 1/3 in the CAR, 
and 0/4 in the ARG.} 
As a caveat related to these rates we note that although the membership of selected stars is quite probable, 
for $\sim$40\% of them no trigonometric parallax information are available, therefore they
cannot be considered as secure members. 
Moreover, especially in the case of the THA, COL, and CAR 
the assignment of some members among these groups is interchangeable. 
High quality data from GAIA (Global Astrometric Interferometer for Astrophysics) 
space observatory will clarify these issues.

Most debris disks around Sun-like moving group members are unresolved and 
their excess has been measured only in a few infrared bands. Generally these data allow us only 
the reliable estimate of characteristic dust temperature and  
 fractional luminosity as fundamental disk parameters. 
Figure~\ref{gstars} shows the fractional luminosities as a function of dust temperatures for 
known debris disks around F8-K7 members of the five groups. As this plot demonstrates 
four of our  
current detections, the disks around HD\,38397, HD\,48370, HD\,160305, and BD-20\,951 
 represent the coldest population within the displayed sample. 

\subsection{Are there potential analogs of the young Solar System's debris disk?}

\begin{figure*} 
\includegraphics[scale=.8,angle=0]{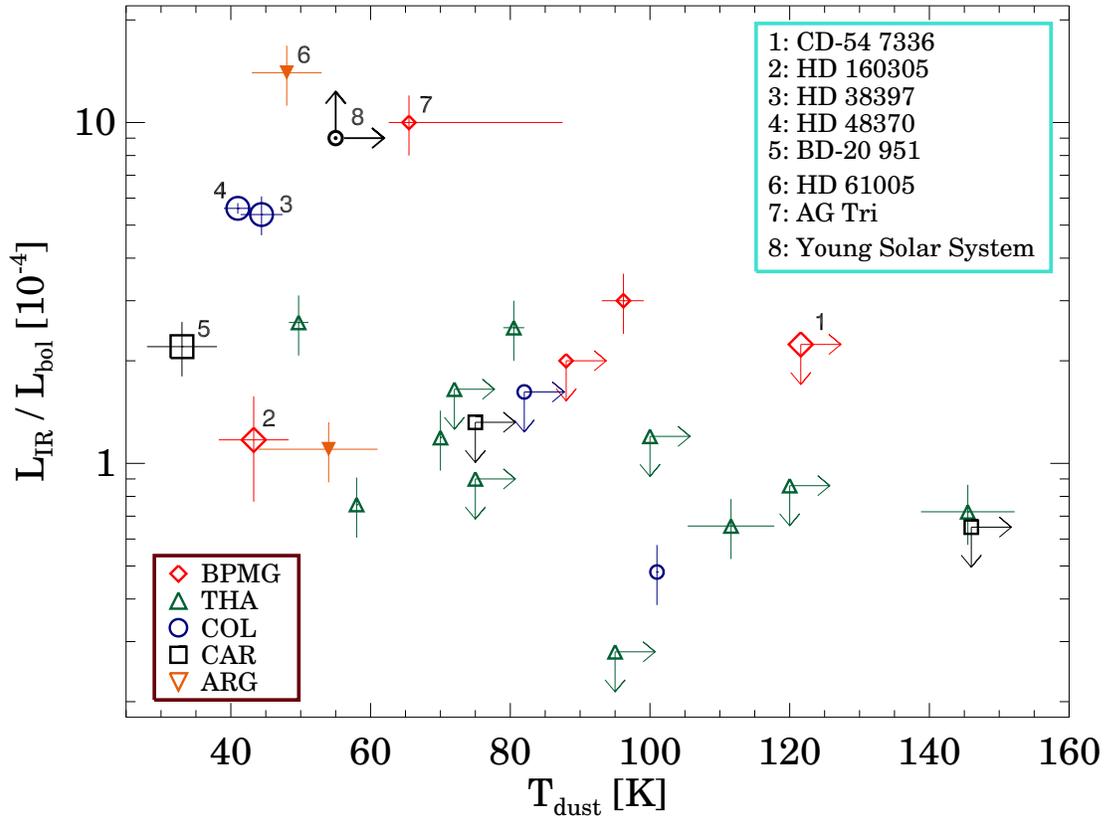}
\caption{ { Fractional luminosities (L$_{\rm IR}$/L$_{\rm bol}$) as a 
function of dust temperature for debris disks encircling {F8-K7-type} members of 
nearby young moving groups (BPMG, THA, COL, CAR, and ARG groups).
Two-temperature systems that may harbor multiple dust belts are displayed with filled symbols. In these  
 cases the colder temperature component has been plotted. Our targets are marked by larger symbols.
For HD\,38397 and HD\,48370 there are hints for warmer components (Sect.~\ref{modelling}), which are not considered here.
The fundamental disk parameters for the other sources were taken from the literature 
\citep{zuckerman2011,donaldson2012,ballering2013,rm2014,chen2014}. 
In those cases where no uncertainty was assigned to the listed fractional luminosities 
we added 20\% error bars. 
{Though AT\,Col, HIP\,84642, and HIP\,30034 were identified as stars with excess at 24{\micron} 
  in \citet{zuckerman2011}, neither their dust temperatures nor fractional luminosities 
  were given in the paper. We derived the missing disk parameters for these sources using the same method as for our target 
  CD-54 7336 (Sect.~\ref{modelling})}. 
}
\label{gstars}
}
\end{figure*}

Comparing to exosolar debris systems, the present disk of the Solar System 
is very tenuous. However, in the early phase of its evolution it may have looked different.
In the first few ten million years after the dispersal of the gas-rich 
primordial disk, terrestrial planets were still forming,
while the giant planets and the Kuiper-belt were in a more compact configuration
within $\sim$35\,au \citep{morbidelli2010}. 
Model calculations for the formation of the largest Kuiper--belt objects and 
binaries indicate that the original belt must have been significantly more massive 
\citep{stern1996,chiang2007,campobagatin2012}.  
According to \citet{booth2009}
this massive predecessor of the Kuiper--belt contained 
 so much dust that the Solar System's debris disk was amongst the brightest 
of such systems at the time. {This calculation is based on the NICE model \citep{gomes2005}, 
which is one possible scenario for the evolution of the early Solar System.} 
To investigate the evolution of Kuiper--belt's dust content, \citet{booth2009} 
utilized four different models. In the simplest one they assumed blackbody grains 
and a single-slope size distribution. 
This model provided the lowest characteristic dust 
temperature and disk fractional luminosity for the whole studied time period. 
In their more realistic models 
a three-phase size distribution was used and besides the simple blackbody grains, amorphous silicates 
and a comet-like grain composition were also evaluated. 
Based on their Figure 11, in the case of comet-like grains, the excess SED peaks at around 40{\micron} 
which would correspond to a characteristic temperature of $\sim$130\,K in a blackbody model. 
It is worth noting that the applied models were focused on the Kuiper--belt and the dust content of the main 
asteroid belt 
was neglected. {Adopting this picture for our further analysis 
in} Figure~\ref{gstars} we plotted the predictions of the simplest model for the given age range (20-40\,Myr),  
 indicating that it is a lower limit both for the dust temperature ($\sim$55\,K) and the fractional 
 luminosity ($\sim$9$\times$10$^{-4}$). 
The derived fractional luminosities for most debris disks around young Sun-like group members as well as
the typical upper limits for stars deemed to have no excess (1--2$\times$10$^{-4}$) 
are significantly lower than the one inferred for the young Solar System. 

{Four debris systems, AG\,Tri (member of the BPMG), HD\,61005 (member of the ARG), HD\,38397 
and HD\,48370 (members of the COL) stand out from the sample based on their 
fractional luminosity.
These disks resemble well the supposed young Kuiper--belt in terms of this parameter, although 
their dust temperature is typically lower. 
While AG\,Tri is a lower mass K6Ve type star that resides in a wide separation binary 
system, the other three stars are  known to be single. Because of this and of their masses 
the last three objects are really good analogs of our Sun: HD\,61005 has a mass of 0.964$\pm$0.025\,M$_\odot$ 
\citep{desidera2011},
while HD\,38397 and HD\,48370 have masses of 1.06$\pm$0.05\,M$_\odot$, and 0.94$\pm$0.05\,M$_\odot$, respectively 
\citep[derived by us based on PARSEC stellar evolutionary models][]{bressan2012}. 
Analyzing the SED of AG\,Tri, both \citet{ballering2013} and \citet{rm2014} claimed 
the presence of a single dust belt around the star.}
The SED of excess emission at HD\,61005 cannot be adequately fitted by a single temperature 
blackbody, but only with a combination of two components, suggesting that the emitting 
dust -- similarly to our Solar System -- is concentrated in multiple 
spatially distinct dust belts around the star \citep[e.g.][]{ricarte2013}.
In contrast with HD\,61005, for HD\,38397 and HD\,48370 no mid-infrared spectrum is available, only the deviation 
of the WISE 22{\micron} photometry from the SED model suggests the presence of warmer grains. 
Without much better coverage in the mid-IR regime, however, one cannot deduce whether these warmer particles 
are located in a separated inner belt. 
The outer disk around HD\,61005 was spatially resolved both in scattered light and at millimeter wavelength
\citep{buenzli2010,olofsson2016,ricarte2013}. Based on these 
observations the disk radius is about 60-70\,au. For HD\,38397 and HD\,48370 we derived disk radii 
of $\sim$90\,au using their spatially resolved 70{\micron} {\sl Herschel} images 
(Sect.~3.3).
{The disk around AG\,Tri has not been resolved yet.}
In summary we can conclude that based on its predicted fractional luminosity 
the young Kuiper--belt would be amongst the brightest debris disks that can be found around 
Sun-like members of nearby young moving groups. Interestingly, in {the resolved} systems with similar dust    
contents the outer dust belt is located $\sim$2-3 farther from their hosts stars than the original 
Kuiper--belt that could be at an average heliocentric distance of 26\,au \citep{booth2009}, thus while 
{HD\,61005, HD\,38397, and HD\,48370} are similar to each other, they cannot be considered 
as perfect analogs of the young Solar System's debris disk.

\subsection{Stirring of planetesimals in the most massive debris disks around young Sun-like stars.} \label{stirring}
{With their ages of 30-40\,Myr, it is a question how the debris disk production in HD\,38397, HD\,48370 and 
HD\,61005 can already be initiated at such large radial distances.}
For effective dust production via collisional erosion of planetesimals, the disk 
needs to be stirred by some mechanism, because without any dynamical excitation 
the collisions would occur at low relative velocities leading to a merge 
between the colliding bodies. 
According to the most commonly invoked self-stirring scenario,  
large, roughly 1000\,km-sized planetesimals embedded in a belt can ignite destructive 
collisions between neighbouring smaller bodies, thereby initiating a collisional 
cascade \citep{kb2004}. However, the gradual build-up of such large bodies -- if they are not present already 
after the disappearance of the protoplanetary disk \citep[e.g. because they have formed 
in streaming instabilities][]{johansen2014} -- via collisional coagulation requires time.  
The larger the distance from the star and the smaller the surface density of the disk, the longer the time 
needed for the formation of 1000\,km-sized bodies, leading to an outward propagating stirring front that is slower 
in a less massive disk. 

\begin{figure} 
\includegraphics[scale=.47,angle=0]{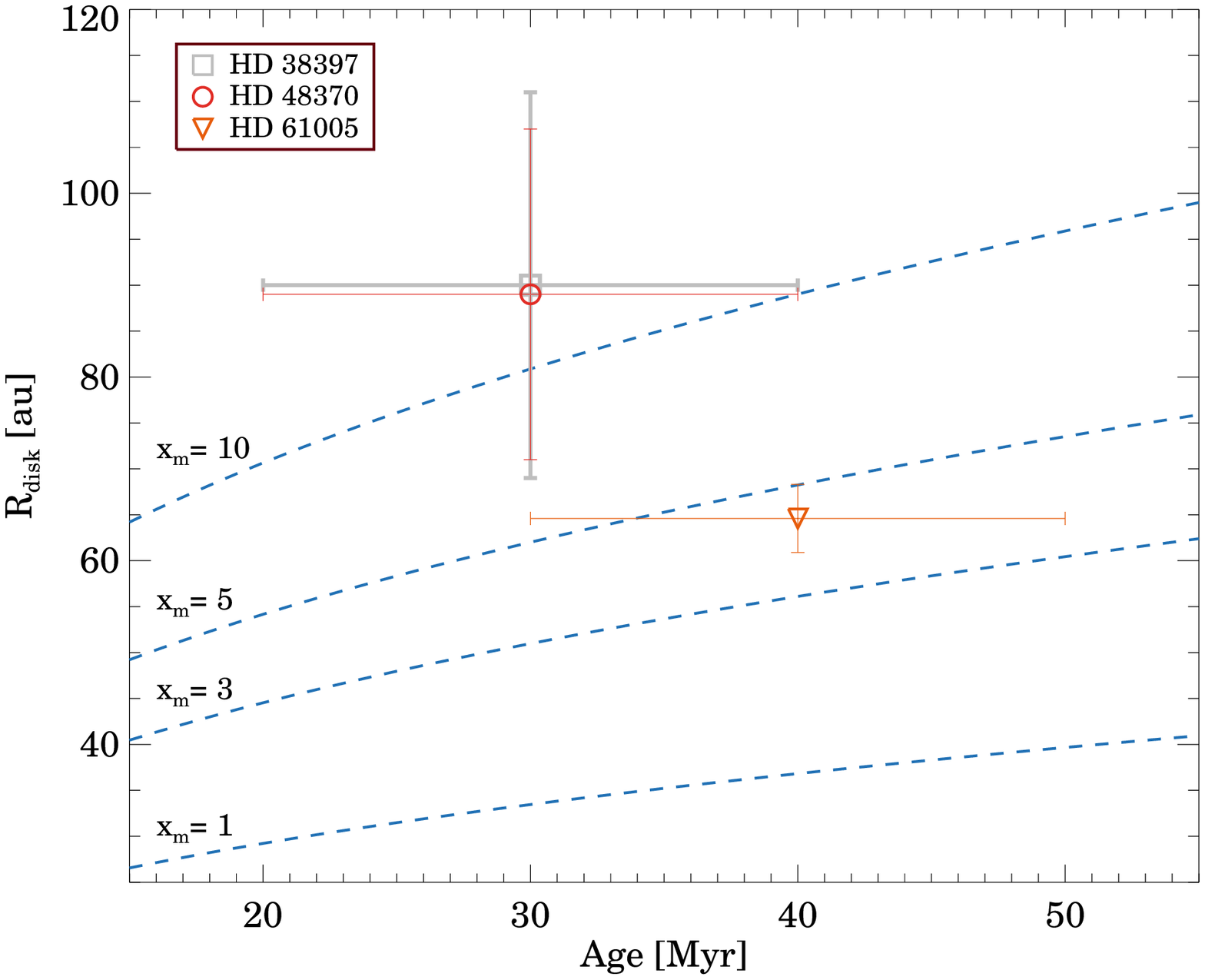}
\caption{ { Disk radii as a function of stellar ages for HD\,38397, HD\,48370, and HD\,61005 
plotted over the predictions of self-stirring models computed for 1.0\,M$_{\odot}$ host stars (blue, dashed lines). 
For the model calculations we used eq.~28. from \citet{kb2008}. According to this formula the timescale 
for the formation of the first 1000\,km planetesimal at a given semimajor axis $a$ around a 1.0\,M$_{\odot}$ star
is $t_{1000} \approx 475 x_m^{-1.15} (\frac{a}{\rm 80 au})^3$\,Myr, where $x_m$ 
is a scaling factor for the reference surface
density of Minimum Mass Solar Nebula. \citet{kb2008} assumed that the initial surface density 
of solid material varies as $\Sigma \propto a^{-3/2}$. 
}
\label{stirringplot}
}
\end{figure}

Using equation~28 from \citet{kb2008} we computed the radius of the stirring front (the radius 
where 1000\,km-size bodies have just been formed) as a function of time 
for disks with different initial surface density ($x_m$ is a scaling factor for the reference surface density 
of Minimum Mass Solar Nebula), assuming a host star with stellar mass of 1\,M$_\odot$. 
In Figure~\ref{stirringplot} we confront the outcomes of this calculations with the disk parameters obtained for the three studied targets (adopting the age of the objects as the time 
available for the formation of large planetesimals). Based on this figure, if the self-stirring mechanism governs 
the dynamical excitation in these systems, then the surface density of their original protoplanetary disks 
 should have been significantly higher than that supposed for the Minimum Mass Solar Nebula.

Alternatively, the motion of planetesimals could be excited by a giant planet via its 
secular perturbation even if the planet is located far from the belt \citep[planetary stirring,][]{wyatt2005b,mustill2009}.
Indeed, in our Solar System both self-stirring and planetary stirring 
 may contribute to the excitation of the Kuiper--belt leading to a high stirring level \citep{matthews2014b}.
If the massive companion orbits closer to the host star than the planetesimals  
 then this mechanism -- similarly to self-stirring -- results in  
 an inside-out process, in which the site of active dust production propagates outward with time. 
   However, for certain planet and disk parameters 
 this process could be even faster than self-stirring, i.e. the collisional cascade
  could be ignited sooner at a given stellocentric distance \citep{mustill2009}. 
By studying spatially resolved images of young debris disks around 
early type stars, \citet{moor2015} identified several systems whose 
radial extent are far too large to be explained 
by self-stirring assuming reasonable
initial disk masses.
For such disks planetary stirring 
is a more feasible scenario. Indeed, in two of these systems, HR\,8799 and HD\,95086, the potential 
planetary mass perturbers have already been discovered via direct imaging \citep{marois2008,rameau2013}.
The debris disks around HD\,38397 and HD\,48370 would also require unusually massive
initial disks in a self-stirring model, thereby these systems are also good candidates for 
planetary stirring. 

Predictions of stirring models are related to the dust producing planetesimals. 
Throughout our previous calculations we assumed that the observed dust grains 
trace the location of planetesimals reliably. However, the spatial distribution 
of debris grains could be different from that of the parent planetesimal belt. Small fragments 
produced in collisions can be moved on more eccentric orbit by the stellar radiation pressure resulting in 
a dust disk that extends outward from the birth ring \citep[e.g.,][]{krivov2010}. 
 In the case of HD\,61005 the disk radius was inferred as the average 
 of size estimates from resolved scattered light and millimeter images \citep{buenzli2010,olofsson2016,ricarte2013}.
 Our original assumption is likely valid for this system, since millimeter emission 
  predominantly traces the distribution of large dust grains with sizes of $>$100{\micron}, 
  that are not affected significantly 
  by radiative forces and thus serve as proxy for the planetesimals.   
At 70{\micron}, where HD\,38397 and HD\,48370 were marginally resolved, we
trace smaller grains that could be sensitive to 
radiation pressure.
Without any information on the grain composition, however, theoretically it is hard to 
judge how much more extended the disk could appear than the ring 
of parent planetesimals at this wavelength.
In \citet{moor2015} we selected some debris disks that were resolved both 
at 70 or 100{\micron} with the {\sl Herschel} and at millimeter wavelengths 
and then compared the pairs of disk sizes derived from the different bands. 
In all cases the outer radii measured in millimeter images were  
larger than the PACS-based average disk radii. For $\epsilon$\,Eri the radial extent 
of the outer dust belt inferred from PACS 70 and 160{\micron} observations \citep{greaves2014} 
is also well consistent within the uncertainties with that derived from recent 
millimeter observation \citep{macgregor2015}. 
These results suggest that this phenomenon may not affect significantly our findings.

\subsection{Disk mass and power law index of dust size distribution at HD\,48370}
The good quality submillimeter/millimeter photometry of HD\,48370 allow us to provide 
estimates both for the disk mass and the power-law index of the dust size distribution ($q$). 
With its mass of $\sim$0.12\,M$_\oplus$ the disk around HD\,48370 is amongst the most massive debris 
disks known to us. By gathering (sub)millimeter data of debris disks with G-type host stars
 from the literature \citep{sheret2004,williams2004,carpenter2005,corder2009,roccatagliatta2009,nilsson2010,
 panic2013,ricarte2013,ricci2015} and estimate their dust masses homogeneously in the same way as in Sect.~\ref{modelling} 
 we found that only HD\,61005 and HD\,107146 harbour similarly massive disks. 

In his pioneering work, \citet{dohnanyi1969} derived $q = 3.5$ as 
the power law index of size distribution for a steady state collisional 
cascade, assuming that both the tensile strength and the velocity 
dispersion of the colliding bodies are independent of their size.
More sophisticated collisional models of \citet{pan2012} and \citet{gaspar2012} 
predict steeper size distributions with $q$ typically larger than 3.6.  
 By fitting the photometric data of HD\,48370 
at $\geq$350{\micron} the power-law index of dust size distribution ($q$) was 
estimated to be 3.25$\pm$0.07 
(Sect.~3.3).
This finding is valid for a grain size interval ranging from 0.1\,mm to a few millimeter
since our observations are primarily sensitive for emission from such dust particles.  
{The derived $q$ value is significantly lower than the theoretically predicted ones. 
Before we compare this value to slopes measured in other debris disks 
and consider the possible physical reasons of this deviation it is worth 
assessing the feasibility of source confusion as an alternative explanation 
of this finding.}   

Since our target was not resolved at millimeter wavelengths, contamination by other sources could not 
be completely excluded, i.e. it is possible that the measured millimeter excesses
are not limited to the disk themselves. However, to be able to result in a 
millimeter slope shallower than that of the disk itself such contaminating sources 
must be comparably bright to the disk and must have a millimeter slope 
shallow enough.


By observing $\alpha$\,Cen\,A/B and
$\epsilon$ Eri 
-- stars with similar spectral types to our {target}
-- with 
radio interferometers, both \citet{liseau2015} and \citet{macgregor2015} found that 
their measured millimeter emission is higher than predicted based on the pure photospheric models.
 The observed excess may be attributed to the stellar chromosphere or 
the corona.
Using 0.87 and 3.1\,mm data \citet{liseau2015} derived spectral indices of 1.62 and 1.61 
for $\alpha$\,Cen\,A and $\alpha$\,Cen\,B, respectively, while \citet{macgregor2015} 
measured a spectral index of $\sim$1.65 between 1.3 and 6.8\,mm in the case of 
$\epsilon$\,Eri, implying that this type of contamination 
could be more significant at longer wavelengths.
 However, the excesses are not very strong  
even at the longest wavelengths, in the cases of $\alpha$\,Cen A/B the total measured 
fluxes at 3.1\,mm are $\sim$2$\times$ higher 
than the predicted photospheric ones, while for
$\epsilon$\,Eri 
a flux enhancement of 3 was derived at 6.8\,mm. Even if we consider 
that HD\,48370 is younger and more active star than the previous 
ones and therefore the flux enhancement might be larger, the level 
of contamination from the star may not have a significant influence on
the measured excess, that is e.g. $\sim$800 $\times$ higher 
than the predicted photospheric flux at 2.14\,mm in the case of HD\,48370.

Large submillimeter/millimeter surveys revealed a massive population of 
luminous high-redshift submillimeter galaxies \citep{blain2002}, 
most of them are site of very intense star formation. 
In order to assess how such background sources can modify 
the derived millimeter slope of our target we took the average SED of ALESS 
sub-millimeter galaxy sample compiled by \citet{dacunha2015} as a template. 
Adopting different redshift values and different flux densities at a reference wavelength of 0.87\,mm, 
we used this template to derive the possible galaxy contamination at 0.35, 0.50, 1.25 and 2.14\,mm. 
Then we subtracted the hypothetical contamination from the measured flux densities of HD\,48370 and  
repeated our fitting process (Sect.~\ref{modelling}) to determine $q$ from the resulted SED.
We found that contamination by galaxies at redshift of $z \gtrsim$3.5 can 
result in the observed slope even if the disk has a $q$ of 3.5 
\citep[correspondingly the prediction of][]{dohnanyi1969}. 
At a redshift of $z=3.5$ such a galaxy must have a 
flux density of 16\,mJy at 0.87\,mm, while at a redshift of 6.0 it should be brighter than 
7\,mJy at 0.87\,mm. In the former case the background galaxy would dominate the combined SED 
at $\geq$0.5\,mm, while in the higher redshift scenario its brightness would be comparable with that 
of the disk at the longest wavelengths of 1.25 and 2.14\,mm. 
Using the formula (eq.~1) of \citet{simpson2015}, we estimated
the number density of submillimeter galaxies 
with flux density of $\geq$7\,mJy at 0.87\,mm to be about 50 sources per square degree. 
Considering the FWHM of IRAM/NIKA beam at 1.25\,mm (12\farcs5) the probability of coincidence with such
 a galaxy is $\sim$5$\times$10$^{-4}$. 
{Thus it is also highly unlikely that such background submillimeter galaxies can explain 
the low $q$ value measured in this debris disk.}

{Indeed, the measured low slope of the power law grain size distribution in HD\,48370 
system is not unique. 
By combining recent VLA 9\,mm and archival ATCA 7\,mm observations 
\citep[taken from][]{ricci2012,ricci2015b} with previous (sub)millimeter data 
for 15 debris disks, \citet{macgregor2016} also found relatively shallow 
slopes with a weighted mean of $q = 3.36\pm$0.02. In their sample there are 
four disks with $q \leq 3.3$.   
Thus, observationally obtained slopes of the solids' size distribution 
found to be typically lower than the values predicted by model calculations 
for collisional cascades of planetesimals. 
\citet{macgregor2016} proposed several physical explanations for the measured 
low $q$ values. They found that adopting a size dependent velocity distribution 
where the velocity increases toward smaller grains, or assuming millimeter, centimeter 
sized grains with strength scaling law similar to 'rubble piles', i.e. held together by gravity 
instead of strength, can result in a grain size distribution consistent 
with the observed ones. In their third scenario, the presence of small grain size cutoff 
due to radiation pressure thought to induce a wavy pattern on the power law grain size distribution \citep[e.g.][]{campo1994}. 
which can affect the determination of $q$, 
providing an alternative explanation for their findings. However, in a  
realistic disk several effects could weaken or smear the expected waviness of 
the size distribution \citep{krivov2006} making this scenario less feasible. 
}     

{Here we note that the} presence of extended regions with low dynamical excitation within the disk 
could also provide a possible explanation for the low measured 
$q$ values. 
According to both the self-stirring and the planetary stirring 
models in an extended planetesimal disk there could be outer regions that are 
not reached by the stirring front and where collisions between bodies 
produce only a lower amount of dust or lead to merging rather than destruction. 
In such dynamically cold environment the bulk of the emitting dust grains 
could be moderately large and radiate like blackbodies \citep{heng2010,krivov2013},  
in regions where the coagulation processes are dominant, the slope 
of the grain size distribution could be very shallow, the value of $q$ could be even 
as low as 2.5 \citep[e.g.,][]{miyake1993}. The observation of the collision dominated 
inner and the coagulation dominated outer parts of a disk without any spatial separation can
result in an averaged, lower $q$ value. 
This scenario might rather be relevant in young systems where the stirring front 
cannot be propagated out to the outer edge of the disk. 
We note that the presence of such a weakly excited extended outer region can also modify 
our conclusions on stirring (Sect.~\ref{stirring}) since in such case the outer edge of a 
dust disk would not necessarily represent the position of the stirring front.




\section{Summary} \label{summary}

Aimed at studying the early evolution of planetesimal
belts and searching for potential analogs of the young Solar System's debris disk 
we studied 31 members of five nearby young moving groups using the
PACS instrument of the {\sl Herschel Space Observatory}. 
None of our targets were observed at far-IR wavelength 
previously.
Most of our selected objects have spectral types between {F8 and K7} thus
can be considered as good analogs of the young Sun. 
Our PACS measurements were complemented by ancillary photometry from the WISE satellite, 
and by continuum submillimeter/millimeter data for one of our sources, HD\,48370.
We identified six stars with IR excess, one of them showing 
excess only in the WISE $W4$ band.
Four of these are new discoveries.
 By assessing the probability of source confusion (particularly due to 
extragalactic objects) we found that the observed excess comes from circumstellar 
debris disks in the majority of these six sources.  
The spectral energy distribution of the targets was 
modeled with a single component modified blackbody representing a narrow dust ring.
In the case of HD\,38397 and HD\,48370 (and to a lesser extent at HD\,160305) 
the excess at 22{\micron} with respect to these models 
hints at a second warmer dust component.    
For HD\,38397 and HD\,48370, the emission is resolved in the 70{\micron} 
PACS images, allowing us to estimate the sizes of these disks.

By combining our new findings with the outcome of other previous infrared surveys
for {those nearby ($<$80\,pc)} Sun-like members of the five selected nearby moving groups 
{that were observed at 70{\micron}}, we 
found that {25 among the 66} systems exhibit some excess emission, the detection rate at 
70{\micron} is {13/66}. 
The latter rate can be considered as a lower limit because 
the stellar photosphere was detected only for the minority of the targets at this wavelength. 
We examined how well these systems resemble the predicted debris 
disk of the young Solar System. {We found that 
concerning their dust content three G-type stars (HD\,61005 and our targets, HD\,38397 and HD\,48370)}
can be considered as analogs of the 
hypothesized ancient Solar System's debris disk, however their outer dust belts 
are located at 2-3 times larger radial distances. 
These three debris systems are
highly interesting by themselves because they represent the highest mass end of the known debris disk 
population around 
young G-type stars.
By evaluating the
feasibility of the self-stirring scenario for these systems we found that 
a reasonable explanation of their current disk structure would require 
a protoplanetary disk that was more massive than the Minimum Mass Solar Nebula.  
Particularly in the case of HD\,38397 and HD\,48370 the required initial surface densities 
raise the possibility that planetary stirring might also contribute to the dynamical 
 excitation of their outer planetesimal belts. This implies that 
 these sources could be promising candidates for direct imaging 
 planet searches. 
Using the measured submillimeter/millimeter data of HD\,48370
we derived a value of 3.25$\pm$0.07 for the slope of the power-law grain size distribution, 
lower than the typical slopes predicted by collisional cascade models.

\acknowledgments
This work was supported by the Momentum grant of the MTA CSFK Lend\"ulet 
Disk Research Group and the Hungarian OTKA grants K104607 and K101393.
AM acknowledges support from the Bolyai
Research Fellowship of the Hungarian Academy of Sciences.
Research of ThH on moving groups is supported by the Heidelberg Collaborative
 Research Center SFB 881 "The Milky Way System". 
{\sl Herschel} is an ESA space observatory with 
science instruments provided by European-led Principal Investigator 
consortia and with important participation from NASA.
Uses observations obtained with the IRAM 30m telescope. IRAM is
supported by INSU/CNRS (France), MPG (Germany), and IGN (Spain).
This publication makes use of data products from the Wide-field
Infrared Survey Explorer, which is a joint project of the University
of California, Los Angeles, and the Jet Propulsion
Laboratory/California Institute of Technology, funded by the National
Aeronautics and Space Administration.  The publication also makes use
of data products from the Two Micron All Sky Survey, which is a joint
project of the University of Massachusetts and the Infrared
Processing and Analysis Center/California Institute of Technology,
funded by the National Aeronautics and Space Administration and the
National Science Foundation. 
PACS has been developed by a consortium of institutes led by MPE (Germany) and
including UVIE (Austria); KU Leuven, CSL, IMEC (Belgium); CEA, LAM (France);
MPIA (Germany); INAF-IFSI/OAA/OAP/OAT, LENS, SISSA (Italy); IAC (Spain). This
development has been supported by the funding agencies BMVIT (Austria),
ESA-PRODEX (Belgium), CEA/CNES (France), DLR (Germany), ASI/INAF (Italy), and
CICYT/MCYT (Spain).
SPIRE has been developed by a consortium of institutes led by Cardiff
University (UK) and including Univ. Lethbridge (Canada); NAOC (China); CEA, LAM
(France); IFSI, Univ. Padua (Italy); IAC (Spain); Stockholm Observatory
(Sweden); Imperial College London, RAL, UCL-MSSL, UKATC, Univ. Sussex (UK); and
Caltech, JPL, NHSC, Univ. Colorado (USA). This development has been supported by
national funding agencies: CSA (Canada); NAOC (China); CEA, CNES, CNRS (France);
ASI (Italy); MCINN (Spain); SNSB (Sweden); STFC, UKSA (UK); and NASA (USA). 
We thank Nicole Pawellek for helpful discussions. 
We also thank the anonymous referee for thoughtful
review.

{\it Facilities:} {\sl Herschel}, {IRAM:30m}.



\floattable
\begin{deluxetable*}{lcccccc}                                                    
\tabletypesize{\footnotesize}                                                           
\tablecaption{Basic stellar properties of the targets. \label{stellarprops}}    
\tablewidth{0pt}                                                                
\tablehead{ \colhead{Target} & \colhead{Spectral type} &                        
\colhead{V} &  \colhead{Distance} & \colhead{Multiplicity} &                    
\colhead{T$_{\rm eff}$} & \colhead{Luminosity} \\                               
\colhead{} & \colhead{} &                                                       
\colhead{(mag)} &  \colhead{(pc)} & \colhead{} &                                
\colhead{(K)} & \colhead{(L$_\odot$)}                                           
 }
\decimalcolnumbers                                                                              
\startdata                                                                      
\multicolumn{7}{c}{BPMG} \\
\hline
                  CD-54 7336 &             K1V &             9.6 &   66$^{\rm T}$ &    N &  5050 &    0.69 \\
                   HD 160305 &      F8/G0V$^*$ &         8.3$^*$ &   72$^{\rm H}$ &    N &  6050 &    1.97 \\
                   HD 168210 &             G5V &             8.9 &   73$^{\rm H}$ &    N &  5650 &    1.43 \\
                 CD-31 16041 &            K7Ve &            11.2 &   51$^{\rm T}$ &    N &  3850 &    0.20 \\
\hline
\multicolumn{7}{c}{THA} \\
\hline
                    CD-78 24 &            K3Ve &            10.2 &   50$^{\rm T}$ &    N &  4550 &    0.25 \\
                    HD 25402 &         G3V$^*$ &         8.4$^*$ &   48$^{\rm H}$ &    N &  5850 &    0.89 \\
             TYC 8083-0455-1 &            K7Ve &            11.5 &   46$^{\rm T}$ &    N &  4000 &    0.11 \\
                  BD-19 1062 &          K3V(e) &            10.6 &   68$^{\rm T}$ &    N &  4650 &    0.31 \\
                  CD-30 2310 &            K4Ve &            11.7 &   65$^{\rm T}$ &    N &  4050 &    0.17 \\
                   CD-86 147 &            G8IV &             9.3 &   60$^{\rm T}$ &    N &  5450 &    0.63 \\
\hline
\multicolumn{7}{c}{COL} \\
\hline
                  CD-36 1785 &            K1Ve &            10.8 &   77$^{\rm T}$ &    N &  4900 &    0.28 \\
               GSC 8077-1788 &            M0Ve &            13.0 &   76$^{\rm T}$ &    Y &  3600 &    0.10 \\
                    HD 31242 &             G5V &             9.9 &   76$^{\rm T}$ &    Y &  5650 &    0.61 \\
                   HD 272836 &          K2V(e) &            10.8 &   76$^{\rm T}$ &    Y &  4800 &    0.34 \\
                    HD 32372 &             G5V &             9.5 &   76$^{\rm H}$ &    N &  5500 &    0.84 \\
                    HD 32309 &         B9V$^*$ &         4.9$^*$ &   61$^{\rm H}$ &    N & 10200 &   41.28 \\
                   HIP 25434 &          G0$^*$ &         9.1$^*$ &   71$^{\rm H}$ &    Y &  6350 &    1.06 \\
                    HD 35996 &   F3/F5IV/V$^*$ &         8.2$^*$ &   71$^{\rm H}$ &    Y &  6700 &    2.03 \\
                   HD 274561 &          K1V(e) &            11.4 &   78$^{\rm T}$ &    N &  4800 &    0.19 \\
                    HD 36329 &             G3V &             9.2 &   71$^{\rm H}$ &    Y &  5800 &    1.73 \\
                    HD 38397 &         G0V$^*$ &         8.1$^*$ &   55$^{\rm H}$ &    N &  6000 &    1.40 \\
                  CD-52 1363 &            G9IV &            10.6 &  106$^{\rm T}$ &    N &  5150 &    0.63 \\
                    HD 48370 &             G8V &             7.9 &   35$^{\rm T}$ &    N &  5600 &    0.72 \\
                    HD 51797 &          K0V(e) &             9.8 &   75$^{\rm T}$ &    N &  5200 &    0.62 \\
 CD-54 4320\tablenotemark{a} &            K5Ve &            10.2 &   44$^{\rm T}$ &    N &  4400 &    0.23 \\
\hline
\multicolumn{7}{c}{CAR} \\
\hline
  BD-20 951\tablenotemark{b} &          K1V(e) &            10.3 &   72$^{\rm T}$ &    Y &  4650 &    0.67 \\
\hline
\multicolumn{7}{c}{ARG} \\
\hline
                      BW Phe &            K3Ve &             9.6 &   41$^{\rm H}$ &    Y &  4600 &    0.52 \\
                  CD-42 2906 &             K1V &            10.6 &   96$^{\rm T}$ &    N &  5100 &    0.49 \\
                  CD-43 3604 &            K4Ve &            10.9 &   79$^{\rm T}$ &    N &  4550 &    0.35 \\
                    HD 67945 &             F0V &             8.1 &   64$^{\rm T}$ &    N &  6950 &    1.92 \\
                  CD-52 9381 &            K6Ve &            10.6 &   29$^{\rm T}$ &    N &  4200 &    0.08 \\
\enddata                                                                                                      
\tablenotetext{a}{\citet{torres2008} proposed this star as a member of                                        
the Carina moving group, but according to the analysis of \citet{malo2013}                                    
it rather belongs to the Columba association.}                                                                
\tablenotetext{b}{\citet{torres2008} assigned BD-20 951 to                                                    
the Tucana-Horologium group, recently, however \citet{elliott2014} proposed this                              
star as a member of the Carina group.}                                                                        
\tablecomments{                                                                                               
 Col.(1): Target name.                                                                                        
 Col.(2): Spectral type, taken from \citet{torres2008} if available,                                          
otherwise (marked by asterisks) {\sl Hipparcos} data were used.                                               
 Col.(3): V band magnitude, taken from \citet{torres2008} if available,                                       
otherwise (marked by asterisks) {\sl Hipparcos} data were used.                                               
 Col.(4): Distance. When {\sl Hipparcos} data were available                                                  
 the distances were computed from the measured trigonometric                                                  
 parallaxes, in other cases the distance data were taken from                                                 
 \citet{torres2008}. For components of wide separation binaries                                               
 (HD\,35996/HIP\,25434 and HD\,31242/GSC\,8077-1788) we                                                       
 listed the average of their estimated distances.                                                             
 Col.(5): Multiplicity. BD-20\,951 and HD\,36329 are SB2 stars                                                
 \citep{torres2008}.                                                                                          
  BW\,Phe and HD 272836 are visual binaries with separations of 0\farcs2                                      
  and 1\farcs52 \citep{torres2008,elliott2015}.                                                               
 GSC\,8077-1788 is a wide separation (18\farcs3) companion of                                                 
  HD\,31242  \citep{torres2008}.                                                                              
HD\,35996 and HIP\,25434 also constitute a wide pair                                                          
with a separation of 12{\arcsec}. \citet{desidera2015} found that HD\,35996                                   
is an SB system that might have an orbital                                                                    
period of a few years.                                                                                        
 Col.(6): Effective temperature. For close binaries with separation                                           
 less than 5{\arcsec},                                                                                        
 we used the combined photometry of the components in the photosphere fitting                                 
 (Sect.~\ref{stellarprop}), hence                                                                             
the derived effective temperatures refer to the combined photospheres.                                        
In the case of BD-20\,951, where \citet{elliott2014} estimated a flux ratio of $\sim$0.25                     
for the two components, the fainter and colder companion may significantly                                    
contribute to the total near-IR fluxes resulting in a lower effective temperature                             
for the whole                                                                                                 
system than that of the primary component.                                                                    
 Col.(7): Luminosity.                                                                                         
}                                                                                                             
\end{deluxetable*}                                                                                             

\floattable
\begin{deluxetable}{lccc} 
\tabletypesize{\footnotesize}                                                     
\tablecaption{Log of Herschel observations. \label{obslog}}                     
\tablewidth{0pt}                                                                
\tablehead{ \colhead{Target} & \colhead{Obs. ID.} &                             
\colhead{Date} &  \colhead{Repetition}                                          
 }  
\decimalcolnumbers                                                                            
\startdata                                                                      
  CD-54 7336 &       1342240321/22\tablenotemark{a} &   2012-03-06 &        3   
          \\
   HD 160305 &       1342241521/22\tablenotemark{a} &   2012-03-16 &        3   
          \\
   HD 168210 &                        1342269686/87 &   2013-04-09 &        2   
          \\
 CD-31 16041 &                        1342267591/92 &   2013-03-14 &        4   
          \\
    CD-78 24 &                        1342269672/73 &   2013-04-09 &        4   
          \\
    HD 25402 &                        1342269646/47 &   2013-04-08 &        2   
          \\
TYC 8083-045 &                        1342269670/71 &   2013-04-09 &        8   
          \\
  BD-19 1062 &                        1342269654/55 &   2013-04-08 &        4   
          \\
  CD-30 2310 &                        1342269256/57 &   2013-04-02 &        8   
          \\
   CD-86 147 &                        1342269674/75 &   2013-04-09 &        4   
          \\
  CD-36 1785 &                        1342269648/49 &   2013-04-08 &        8   
          \\
GSC 8077-178 &                        1342269666/67 &   2013-04-09 &        4   
          \\
    HD 31242 &                        1342269666/67 &   2013-04-09 &        4   
          \\
   HD 272836 &                        1342269668/69 &   2013-04-09 &        4   
          \\
    HD 32372 &                        1342269664/65 &   2013-04-09 &        4   
          \\
    HD 32309 &                        1342269652/53 &   2013-04-08 &        2   
          \\
   HIP 25434 &                        1342269662/63 &   2013-04-09 &        2   
          \\
    HD 35996 &                        1342269662/63 &   2013-04-09 &        2   
          \\
   HD 274561 &                        1342269252/53 &   2013-04-02 &        8   
          \\
    HD 36329 &                        1342269254/55 &   2013-04-02 &        2   
          \\
    HD 38397 &       1342242088/89\tablenotemark{a} &   2012-03-20 &        3   
          \\
  CD-52 1363 &                        1342269250/51 &   2013-04-02 &        4   
          \\
    HD 48370 &                        1342269656/57 &   2013-04-08 &        2   
          \\
             &       1342250790/91\tablenotemark{b} &   2012-09-09 &   \ldots \\
             &       1342253429/30\tablenotemark{b} &   2012-10-15 &   \ldots \\
    HD 51797 &                        1342269660/61 &   2013-04-08 &        4   
          \\
  CD-54 4320 &                        1342265607/08 &   2013-02-21 &        4   
          \\
   BD-20 951 &                        1342269650/51 &   2013-04-08 &        4   
          \\
      BW Phe &                        1342270953/54 &   2013-04-27 &        2   
          \\
  CD-42 2906 &                        1342265518/19 &   2013-02-20 &        8   
          \\
  CD-43 3604 &                        1342269658/59 &   2013-04-08 &        4   
          \\
    HD 67945 &                        1342269779/80 &   2013-04-10 &        2   
          \\
  CD-52 9381 &                        1342267766/67 &   2013-03-17 &        4   
          \\
\enddata                                                                        
\tablenotetext{a}{PACS mini scanmap observations taken from                     
the Herschel Archive (OT1\_dpadgett\_1 programme, PI: D. Padgett).}             
\tablenotetext{b}{SPIRE/PACS parallel maps.}                                    
\tablecomments{                                                                 
 Col.(1): Target name.                                                          
 Col.(2): Observation identifier.                                               
 Col.(3): Date of the observation.                                              
 Col.(4): Repetition factor of PACS mini scanmap observations.                  
}                                                                               
\end{deluxetable}                                                               


\floattable
\begin{deluxetable*}{lrcrrcrrcrc} 
\setlength{\tabcolsep}{6pt}                                            
\tabletypesize{\footnotesize}                                                           
\tablecaption{Photometric data. \label{phottable}}                              
\tablewidth{0pt}                                                                
\tablehead{ \colhead{Target} & \multicolumn{3}{c}{22.088$\mu$m} &               
\multicolumn{3}{c}{70$\mu$m}  &                                                 
            \multicolumn{3}{c}{160$\mu$m} & \colhead{  }\\                      
\cline{2-4} \cline{5-7} \cline{8-10}\\                                          
\colhead{}  & \colhead{F$_{\rm W4}$}  & \colhead{P$_{\rm W4}$}  &               
	      \colhead{$\chi_{\rm W4}$} & \colhead{F$_{\rm 70}$ [mJy]}  &              
	      \colhead{P$_{\rm 70}$ [mJy]}  &                                          
\colhead{$\chi_{\rm 70}$} & \colhead{F$_{\rm 160}$ [mJy]}  &                    
\colhead{P$_{\rm 160}$ [mJy]}  &                                                
 \colhead{$\chi_{\rm 160}$} &                                                   
 \colhead{ $L_{\rm IR} / L_{\rm bol}$ [10$^{-4}$] }                             
 }  
\decimalcolnumbers                                                                            
\startdata                                                                      
        CD-54 7336 &         14.4$\pm$1.2 &     9.4 &   $+$3.8 &          8.9$\pm$3.2 &     0.9 &   \ldots &         3.9$\pm$18.1 &  0.18 &   \ldots &          $<$2.2 \\
         HD 160305 &         17.7$\pm$1.5 &    13.0 &   $+$2.8 &         26.5$\pm$4.3 &     1.3 &   $+$5.9 &         26.5$\pm$8.5 &    0.24 &   $+$3.1 &     $1.2\pm0.4$ \\
         HD 168210 &         14.6$\pm$1.7 &    11.6 &   $+$1.7 &                           4.5$\pm$3.4 &     1.1 &   \ldots &        46.0$\pm$12.4\tablenotemark{a} &    0.21 &   \ldots &          $<$0.8 \\
       CD-31 16041 &          9.2$\pm$1.3 &    10.8 &   $-$1.2 &       $-$0.8$\pm$2.3 &     1.1 &   \ldots &          3.0$\pm$8.0 &    0.20 &   \ldots &          $<$1.8 \\
          CD-78 24 &          8.2$\pm$0.9 &     8.3 &   $-$0.1 &       $-$2.4$\pm$2.4 &     0.8 &   \ldots &          0.6$\pm$8.7 &    0.16 &   \ldots &          $<$1.6 \\
          HD 25402 &         16.6$\pm$1.2 &    14.5 &   $+$1.5 &          6.8$\pm$2.7 &     1.4 &   \ldots &       $-$9.9$\pm$9.6 &    0.27 &   \ldots &          $<$0.5 \\
   TYC 8083-0455-1 &          7.8$\pm$0.7 &     6.7 &   $+$1.3 &          0.1$\pm$2.0 &     0.7 &   \ldots &          0.7$\pm$5.5 &    0.13 &   \ldots &          $<$2.3 \\
        BD-19 1062 &          5.7$\pm$0.9 &     5.2 &   $+$0.5 &          6.0$\pm$2.1 &     0.5 &   \ldots &          1.8$\pm$7.2 &    0.10 &   \ldots &          $<$2.0 \\
        CD-30 2310 &          4.7$\pm$0.8 &     4.7 &   $+$0.0 &       $-$1.9$\pm$2.0 &     0.5 &   \ldots &          0.2$\pm$9.4 &    0.09 &   \ldots &          $<$3.2 \\
         CD-86 147 &          8.5$\pm$0.9 &     8.2 &   $+$0.3 &          1.8$\pm$3.3 &     0.8 &   \ldots &      $-$14.0$\pm$8.2 &    0.15 &   \ldots &          $<$1.2 \\
        CD-36 1785 &          3.0$\pm$0.7 &     3.0 &   $+$0.0 &       $-$1.3$\pm$1.8 &     0.3 &   \ldots &      $-$12.6$\pm$5.4 &    0.06 &   \ldots &          $<$3.9 \\
     GSC 8077-1788 &          3.2$\pm$0.6 &     3.2 &   $+$0.1 &       $-$1.1$\pm$2.0 &     0.3 &   \ldots &          4.5$\pm$5.6 &    0.06 &   \ldots &          $<$7.0 \\
          HD 31242 &          5.8$\pm$0.7 &     4.5 &   $+$1.7 &          1.6$\pm$2.0 &     0.4 &   \ldots &       $-$0.8$\pm$6.4 &    0.08 &   \ldots &          $<$1.2 \\
         HD 272836 &          4.7$\pm$0.7 &     4.2 &   $+$0.7 &          3.0$\pm$1.8 &     0.4 &   \ldots &          2.9$\pm$3.9 &    0.08 &   \ldots &          $<$1.9 \\
          HD 32372 &          8.4$\pm$0.8 &     6.7 &   $+$2.0 &          7.3$\pm$2.4 &     0.7 &   $+$2.8 &     $-$15.9$\pm$10.4 &    0.12 &   \ldots &          $<$1.0 \\
          HD 32309 &         87.5$\pm$5.5 &    77.9 &   $+$1.4 &          5.9$\pm$1.5 &     7.6 &   $-$1.1 &          9.7$\pm$9.1 &    1.42 &   \ldots &        $<$0.009 \\
         HIP 25434 &          6.4$\pm$0.7 &     6.3 &   $+$0.1 &          1.4$\pm$2.2 &     0.6 &   \ldots &       $-$6.4$\pm$7.3 &    0.12 &   \ldots &          $<$0.7 \\
          HD 35996 &         16.9$\pm$1.2 &    10.4 &   $+$4.8 &          8.7$\pm$1.9 &     1.0 &   $+$4.0 &       $-$2.4$\pm$7.7 &    0.19 &   \ldots &     $1.0\pm0.3$ \\
         HD 274561 &          1.7$\pm$0.6 &     2.2 &   $-$0.7 &          1.2$\pm$1.6 &     0.2 &   \ldots &          5.2$\pm$4.7 &    0.04 &   \ldots &          $<$3.3 \\
          HD 36329 &         14.9$\pm$1.2 &    13.6 &   $+$0.9 &          6.1$\pm$2.8 &     1.3 &   \ldots &          0.5$\pm$5.0 &    0.25 &   \ldots &          $<$0.5 \\
          HD 38397 &         23.1$\pm$1.6 &    16.3 &   $+$3.8 &       141.4$\pm$11.1 &     1.6 &  $+$12.6 &       144.7$\pm$17.2 &    0.30 &   $+$8.4 &     $5.4\pm0.7$\tablenotemark{c} \\
        CD-52 1363 &          4.2$\pm$0.6 &     3.1 &   $+$1.8 &          4.0$\pm$2.4 &     0.3 &   \ldots &       $-$3.4$\pm$8.7 &    0.06 &   \ldots &          $<$2.7 \\
          HD 48370 &         36.5$\pm$2.5 &    25.6 &   $+$3.8 &       176.6$\pm$14.6 &     2.5 &  $+$12.1 &       220.3$\pm$20.7 &    0.47 &  $+$11.2 &      $5.6\pm0.2$\tablenotemark{c} \\
                   &                      &         &          &       156.7$\pm$20.0\tablenotemark{b} &         &   $+$7.7 &       216.4$\pm$28.8\tablenotemark{b} &         &   $+$7.4 &                 \\
          HD 51797 &          8.2$\pm$0.9 &     6.1 &   $+$2.4 &          6.4$\pm$2.1 &     0.6 &   $+$2.7 &       $-$8.9$\pm$6.2 &    0.11 &   \ldots &          $<$1.2 \\
        CD-54 4320 &         13.3$\pm$1.1 &    11.0 &   $+$2.0 &          0.8$\pm$2.2 &     1.1 &   \ldots &         16.3$\pm$9.6 &    0.21 &   \ldots &          $<$1.1 \\
         BD-20 951 &         11.7$\pm$1.0 &    10.2 &   $+$1.4 &         12.5$\pm$2.3 &     1.0 &   $+$5.0 &         24.6$\pm$6.2 &    0.19 &   $+$3.9 &     $2.2\pm0.4$ \\
            BW Phe &         27.2$\pm$1.8 &    24.8 &   $+$1.0 &          6.9$\pm$2.3 &     2.5 &   $+$1.9 &        25.5$\pm$10.3 &    0.47 &   \ldots &          $<$0.5 \\
        CD-42 2906 &          3.8$\pm$0.7 &     3.1 &   $+$1.0 &       $-$2.7$\pm$2.0 &     0.3 &   \ldots &     $-$12.9$\pm$11.6 &    0.06 &   \ldots &          $<$2.4 \\
        CD-43 3604 &          5.7$\pm$0.8 &     4.6 &   $+$1.3 &          2.2$\pm$2.4 &     0.5 &   \ldots &         12.3$\pm$9.9 &    0.09 &   \ldots &          $<$2.7 \\
          HD 67945 &         10.2$\pm$1.1 &    10.8 &   $-$0.5 &          1.7$\pm$2.8 &     1.1 &   \ldots &       $-$2.9$\pm$6.0 &    0.20 &   \ldots &          $<$0.4 \\
        CD-52 9381 &         11.7$\pm$1.2 &    10.5 &   $+$0.9 &       $-$2.2$\pm$2.0 &     1.1 &   \ldots &          1.0$\pm$8.8 &    0.20 &   \ldots &          $<$1.3 \\
\enddata                                                                        
\tablenotetext{a}{Extended nebulosity                                           
contaminated the area of the aperture.}                                         
\tablenotetext{b}{Photometry from SPIRE/PACS parallel                           
maps.}
\tablenotetext{c}{Note that single temperature models do not reproduce WISE 22{\micron} excess (see Sect.~\ref{modelling}). 
By adopting an additional warmer component, the two-temperature fit results in a higher total fractional luminosity.}                                                                         
\tablecomments{                                                                 
 Col.(1): Target name.                                                          
 Col.(2-10): Measured and predicted flux densities with their                   
 uncertainties and the significance of                                          
 the excesses ($\chi_{\rm band}$) in WISE W4 band and in PACS 70 and            
 160{\micron}                                                                   
 bands. The                                                                     
 quoted uncertainties include the calibration uncertainties as well.            
 No color correction was applied.                                               
 $\chi_{\rm 70}$ and $\chi_{\rm 160}$ were quoted only those cases where the    
 target was detected in the specific band.                                      
}                                                                               
\end{deluxetable*}

\floattable
\begin{deluxetable}{ccccc}                                                                                              
\tabletypesize{\scriptsize}                                                                                             
\tablecaption{Measured and predicted fluxes for HD\,48370 \label{phottable2}}                                           
\tablewidth{0pt}                                                                                                        
\tablehead{ \colhead{$\lambda$} & \colhead{Instr.} & \colhead{Meas. flux$^{a}$} &                                       
 \colhead{Pred. flux} &  \colhead{Reference} \\                                                                         
 \colhead{({\micron})} & \colhead{} & \colhead{(mJy)} &                                                                 
 \colhead{(mJy)} &  \colhead{}                                                                                          
} 
\decimalcolnumbers                                                                                                        
\startdata                                                                                                              
        3.35 &               WISE &         955.6$\pm$45.8 &    957.1 &   \citet{wright2010} \\
       11.56 &               WISE &           91.9$\pm$4.3 &     91.4 &   \citet{wright2010} \\
       22.09 &               WISE &           36.5$\pm$2.5 &     25.6 &   \citet{wright2010} \\
   70.00$^b$ &               PACS &         173.0$\pm$14.0 &      2.5 &            this work \\
      160.00 &               PACS &         219.4$\pm$19.5 &     0.47 &            this work \\
      250.00 &              SPIRE &         148.5$\pm$12.1 &     0.19 &            this work \\
      350.00 &              SPIRE &         103.4$\pm$11.6 &     0.10 &            this work \\
      500.00 &              SPIRE &          54.5$\pm$10.5 &     0.05 &            this work \\
     1250.00 &               NIKA &            7.3$\pm$1.5 &    0.007 &            this work \\
     2140.00 &               NIKA &            2.0$\pm$0.3 &    0.002 &            this work \\
\enddata                                                                        
\tablenotetext{a}{The quoted flux densities are not color corrected.}           
\tablenotetext{b}{The emission is resolved at this wavelength.}                 
\end{deluxetable}                                                               

\floattable
\tabletypesize{\scriptsize}
\begin{deluxetable}{lcccc}                                                                  
\tablecaption{Disk properties from a single temperature modified blackbody fit to the observed excess 
 \label{diskparams}}                                                                        
\tablewidth{0pt}                                                                            
\tablehead{\colhead{Target}  &  \colhead{$\rm T_{\rm dust}$ (K)} & \colhead{$\beta$}        
&  \colhead{$\rm f_{\rm dust}$ ($\rm 10^{-4}$)} & \colhead{$\rm R_{\rm dust}$ (au)}         
}
\decimalcolnumbers                                                                                           
\startdata  
  {CD-54 7336}   & $>$122   & 0.65   & {$<$2.2}  & $<$4.3 \\
  { HD\,160305}   & 43$\pm$5   & 0.65   & 1.2$\pm$0.4  & 58$\pm$13 \\                       
  { HD\,35996}   & 135$\pm$17   & 0.65   &1.0$\pm$0.3  & 6$\pm$2 \\ 
  { HD\,38397}   &  44$\pm$3   & 0.65   & 5.4$\pm$0.7  & 46$\pm$7 \\
  { BD\,48370}   & 41$\pm$2   & 0.26$\pm$0.08   & 5.6$\pm$0.2  & 39$\pm$5 \\
  { BD-20\,951}  & 33$\pm$5   & 0.65   & 2.2$\pm$0.4  & 55$\pm$20 \\ 
 \enddata                                                                                                            \tablecomments{ 
 Col.(1): Target name.                                                                      
 Col.(2): Characteristic disk temperature.                                                  
 Col.(3): Power law index of the emissivity ($\beta$). Except for HD\,48370
 the value of $\beta$ was fixed to 0.65.                                                    
 Col.(4): Fractional dust luminosity, $f_{\rm dust} = \frac{L_{\rm dust}}{L_{\rm bol}}$.    
 Col.(5): Blackbody radius of the disk.                                                     
}                                                                                           
\end{deluxetable} 



\floattable
\begin{deluxetable}{lcccc}                                                      
\tabletypesize{\scriptsize}                                                           
\tablecaption{References for infrared excess detections \label{excesstable}}    
\tablewidth{0pt}                                                                
\tablehead{ \colhead{Target} & \colhead{Spectral type} &                        
\colhead{Excess?} &  \colhead{Excess at 70{\micron}?} & \colhead{Ref.}          
 }
\decimalcolnumbers                                                                              
\startdata                                                                      
\multicolumn{5}{c}{BPMG} \\
\hline
                   HIP 10679 &             G2V &    Y &    Y &   6,7 \\
                      AG Tri &            K6Ve &    Y &    Y &   6,7 \\
                   BD+05 378 &            K6Ve &    N &    N &   6,7 \\
                      AO Men &            K4Ve &    N &    N &   6,7 \\
                    V343 Nor &             K0V &    N &    N &   6,7 \\
                    V824 Ara &            G7IV &    N &    N &     6 \\
                  CD-64 1208 &            K5Ve &    N &    N &   6,7 \\
                      PZ Tel &            G9IV &    N &    N &   6,7 \\
                      AZ Cap &            K6Ve &    Y &    N &     5 \\
\hline
\multicolumn{5}{c}{THA} \\
\hline
                      HD 105 &             G0V &    Y &    Y &     4 \\
                      HD 987 &             G8V &    Y &    Y &     8 \\
                     HD 1466 &             F8V &    Y &    Y &     4 \\
                     HD 3221 &            K4Ve &    N &    N &     4 \\
                     HD 8558 &             G7V &    Y &    N &     8 \\
                      CC Phe &             K1V &    N &    N &     8 \\
                      DK Cet &             G4V &    Y &    Y &     4 \\
   HD 13183\tablenotemark{a} &             G7V &   Y? &    N &     8 \\
                   CD-60 416 &            K5Ve &    N &    N &     8 \\
                   CD-53 544 &            K6Ve &    N &    N &     8 \\
                   CD-58 553 &            K5Ve &    N &    N &     8 \\
 CD-46 1064\tablenotemark{a} &            K3Ve &   Y? &    N &     8 \\
                    HD 22705 &             G2V &    Y &    N &     8 \\
                    HD 29615 &             G3V &    N &    N &     8 \\
                   HIP 84642 &             G8V &    Y &    N &     8 \\
                      BO Mic &            K4Ve &    Y &    N &    1,5\\
                   HD 202917 &             G7V &    Y &    Y &     4 \\
                  HIP 105404 &             G9V &    Y &    N &     8 \\
                  HIP 108422 &            G9IV &    Y &    N &     8 \\
                  HD 222259B &            K3Ve &    N &    N &     8 \\
                      DS Tuc &             G6V &    N &    N &     8 \\
\hline
\multicolumn{5}{c}{COL} \\
\hline
                   BD-15 705 &           K3(e) &    N &    N &     8 \\
                   BD+08 742 &             G5: &    N &    N &     2 \\
                    HD 36869 &             G3V &    N &    N &     8 \\
                      AT Col &            K1Ve &    Y &    N &     8 \\
                   V1358 Ori &             F9V &    Y &    N &     1 \\
\hline
\multicolumn{5}{c}{CAR} \\
\hline
                   HIP 30034 &          K1V(e) &    Y &    N &   4,8 \\
                    HD 49855 &             G6V &    Y &    N &     8 \\
                    HD 55279 &             K2V &    N &    N &     4 \\
\hline
\multicolumn{5}{c}{ARG} \\
\hline
                    HD 61005 &           G8V k &    Y &    Y &     3 \\
                    HD 84075 &              G1 &    Y &    Y &     3 \\
                   CD-74 673 &            K3Ve &    N &    N &     8 \\
\enddata                                                                                                      
\tablenotetext{a}{For these objects \citet{zuckerman2011} reported                                            
weak excess at 24{\micron} but claimed that the identification                                                
of the excesses is uncertain. Therefore in our analysis in Sect. 4.2                                          
we will not consider them as infrared excess stars.}                                                          
\tablecomments{                                                                                               
 Col.(1): Target name.                                                                                        
 Col.(2): Spectral type.                                                                                      
 Col.(3): Presence of infrared excess.                                                                        
 Col.(4): Presence of infrared excess at 70{\micron}.                                                         
 Col.(5): References for information on IR excess.                                                            
 1 -- \citet{ballering2013}; 2 -- \citet{carpenter2009};                                                      
 3 -- \citet{chen2014}; 4 -- \citet{donaldson2012};                                                           
 5 -- \citet{plavchan2009}; 6 -- \citet{rebull2008};                                                          
 7 -- \citet{rm2014}; 8 -- \citet{zuckerman2011}.                                                             
}                                                                                                             
\end{deluxetable}



\end{document}